\documentclass[fleqn,usenatbib]{mnras}

\usepackage[T1]{fontenc}

\DeclareRobustCommand{\VAN}[3]{#2}
\let\VANthebibliography\thebibliography
\def\thebibliography{\DeclareRobustCommand{\VAN}[3]{##3}\VANthebibliography}

\usepackage{graphicx}	
\usepackage{amsmath}	
\usepackage{amssymb}	
\usepackage{newtxtext,newtxmath}
\usepackage[dvipsnames]{xcolor}
\usepackage{multirow}
\usepackage{xcolor}
\usepackage{{booktabs}}
\usepackage{{longtable}}
\usepackage{pdflscape}
\usepackage{caption}
\usepackage{subcaption}

\newcommand{\ero}{\textit{eROSITA}}

\newcommand{\ergps}{erg\,s$^{-1}$}
\newcommand{\ergpspcm}{erg \,s$^{-1}$\,cm$^{-2}$}

\newcommand{\stmass}{M_{\rm *}}
\newcommand{\msun}{M_\odot}
\newcommand{\lxlxrbrat}{\frac{L_{\rm X, obs}}{L_{\rm X, XRB}}}
\newcommand{\dd}{\mathrm{d}}

\title[SRG/eROSITA catalogue of active dwarf galaxies]{SRG/eROSITA catalogue of X-ray active SDSS dwarf galaxies}

\author[Bykov, Gilfanov \& Sunyaev]{
S.D. Bykov$^{1,2}$\thanks{E-mail:sergei.d.bykov@gmail.com}, M.R. Gilfanov$^{3,2}$
 \& R.A. Sunyaev$^{3,2}$\\
$^{1}$Kazan Federal University, 18 Kremlyovskaya Street, Kazan, Russia\\
$^{2}$Max Planck Institute for Astrophysics, Karl-Schwarzschild-Str 1, Garching b. München D-85741, Germany\\
$^{3}$Space Research Institute, Russian Academy of Sciences, Profsoyuznaya 84/32, 117997 Moscow, Russia\\
}

\date{Accepted XXX. Received YYY; in original form ZZZ}

\pubyear{2023}

\begin{document}
\label{firstpage}
\pagerange{\pageref{firstpage}--\pageref{lastpage}}
\maketitle

\begin{abstract}
We present a sample of 99 dwarf galaxies ($M_*<10^{9.5} M_\odot$) with X-ray activity in their central regions. The sample was obtained from a match of the SRG/eROSITA X-ray catalogue in the Eastern Galactic hemisphere with the MPA-JHU SDSS catalogue. The obtained matches were cleaned rigorously with the help of external optical catalogues to increase the purity of the sample. This work is the largest study of this kind -- X-ray activity in  $\approx 85$ per cent of matched dwarfs was not reported before. The majority of  X-ray active dwarfs are identified as star-forming galaxies. However, the X-ray luminosity of 82 objects cannot be  explained by the collective emission of X-ray binaries, rendering them strong candidates for dwarf galaxies with an active accreting black hole in their centre. We find that the fraction of AGN among dwarf galaxies drops from $\sim 2\cdot 10^{-2}$ at $L_X\sim 10^{39}$ erg/s to $\sim (2-4)\cdot 10^{-4}$ at $L_X\sim 10^{41}$ erg/s and increases with the stellar mass of the host galaxy. We serendipitously discovered sources with unexpected properties. We report on a tidal disruption event (TDE) candidate in a dwarf galaxy, a massive black hole in a dwarf galaxy with a soft thermal spectrum, a luminous dwarf galaxy with an obscured X-ray spectrum and a few other peculiar sources.  We found three Ultra-luminous X-ray source (ULX) candidates and a sample of X-ray bright galaxy pairs, in four of which both members shine in X-rays.

\end{abstract}




\section{Introduction}
\label{sect:intro}

It is known that virtually every massive galaxy hosts a supermassive black hole (SMBH) in its core \citep{Kormendy2013}. The observed correlations between SMBH mass and galaxy bulge properties (such as velocity dispersion, luminosity, and mass) led to the belief that the evolution of SMBHs is intimately related to the evolution of host galaxies \citep{Ferrarese2000, McConnell2013}. 
In less massive galaxies (stellar mass $\stmass<10^{9.5} \msun$, so-called dwarf galaxies) the demography of central black hole population is not so explicit because of the difficulties in the detection  of nuclear activity \citep{Greene2020, Reines2022}.
The role of a central black hole in dwarf galaxies might be the key to the solution to several cosmological and galaxy evolution problems. Examples include AGN feedback, outflows and suppression of star formation in dwarfs \citep{Calabro2017, Barai2019, Manzano-King2019, Ferre-Mateu2021}, core-cusp problem and the reionisation of the Universe \citep{Silk2017}. 
Finally, the importance of dwarfs is emphasised simply by their sheer abundance -  it is the most abundant type of galaxy in the Universe. 

Additional consideration comes from the fact that the formation and evolution of SMBH is not yet understood, and theories are challenged by the presence of very massive quasars within the first Gyr after the Big Bang  \citep{Volonteri2010, Woods2019, Lusso2023}. Possible scenarios for the birth of such SMBHs involve the seed black holes formed as the remnants of Population III stars \citep{Madau2001}, heavier seed such as the direct collapse of gas clouds \citep{Loeb1994}, or other models \citep{Mezcua2017}. The less massive black holes at the centres of dwarfs (which may be of intermediate mass, IMBH) still have imprints of the precursor BHs formation and evolution \citep{Volonteri2010}. Moreover, dwarf galaxies can  help the understanding of the seeding mechanism because they have relatively poor merger and accretion history, thus their BHs are the closest analogues of the seed BHs \citep{Mezcua2017, Zubovas2019, Greene2020, Reines2022, Burke2023}. 

There are several ways to find black holes in dwarf (and massive) galaxies. The standard way is the analysis of optical emission line spectra for the signatures of an active (i.e. accreting) galactic nucleus \citep{Reines2013, Chilingarian2018, Molina2021}. Other possibilities are optical variability \citep{Baldassare2020, Ward2022}, radio \citep{Reines2011, Reines2020}, mid-infrared  \citep{Sartori2015, Marleau2017} and X-ray \citep{Birchall2022, Zou2023} selections. All methods are prone to selection effects and contamination by non-AGN-related processes mimicking AGN signatures, such as supernovae and their remnants, star formation activity, and tidal disruption events (TDE) among others. The mentioned methods allow selection only of active black holes, whereas finding quiescent (dormant) black holes is much more difficult and addressed with dynamical methods, TDE detection or gravitational waves, see  \citet{Reines2022}. 

X-ray emission is a reliable indicator of accretion onto the central black hole if it is not confused with a TDE, aggregated X-ray binaries emission or a background AGN. X-ray selection allows picking up objects accreting at lower Eddington ratios compared to other search methods \citep{Birchall2022}.  The usual approach is to take a stellar-mass selected sample of galaxies and search for X-ray emission. X-ray data from Chandra or XMM-Newton are often employed \citep{Pardo2016, Lemons2015, Baldassare2017,  Birchall2020, Mezcua2023}, and recently -- SRG/eROSITA \citep{Latimer2021}. The largest X-ray selected samples come from XMM-SERVS (73 candidates from \citealt{Zou2023}), 3XMM (61 candidates from \citealt{Birchall2020}) and Chandra COSMOS (40 candidates from \citealt{Mezcua2018}). Small-area X-ray surveys (COSMOS, XMM-SERVS) allow finding more distant and faint dwarfs AGN (up to redshift $z\sim2.5$), whereas wide-angle surveys (3XMM) pick up the local population ($z\lesssim0.3$) and more rare and luminous objects. Hard X-ray (>10 keV) selection proposes a method less prone to absorption biases and ULX/X-ray binaries contamination, but limited by shallow flux limits \citep[e.g.][]{Mereminskiy2023}.  Another example of X-ray data helping to find active dwarfs comes from the work of \citet{Chilingarian2018}, who studied the properties of optical spectra to search for the IMBH candidates and only then invoked X-ray data to find 10 bona fide accreting IMBHs.  If the contribution from X-ray binaries is subtracted properly and the selection effects are taken into account, the occupation fraction of central BHs can be studied in detail with its dependencies on luminosity, stellar mass and redshift \citep{Mezcua2018, Birchall2020, Zou2023}.   

This paper focuses on the search for active nuclei in dwarf galaxies with the help of X-ray data from the SRG/eROSITA all-sky survey, eRASS. In section \ref{sect:data} we describe the data used to construct the sample: catalogue of dwarf galaxies from SDSS (subsect. \ref{sect:data:galaxy}) and eROSITA data (\ref{sect:data:xray}). Section \ref{sect:catalog} explains the process of X-ray active dwarf catalogue construction and cleaning steps (\ref{sect:catalog:cross-match}), analysis of X-ray data (\ref{sect:catalog:xray}), and the estimation of the contamination from X-ray binaries and hot interstellar gas (\ref{sect:catalog:xrb}).  We present the catalogue properties and the discussion of individual sources in sect.  \ref{sect:results} and \ref{sect:indiv} respectively. In sect. \ref{sect:conclusion} we conclude.

We use decimal logarithms in the paper and assume fiducial cosmological parameters, $H_0=70$ km s$^{-1}$ Mpc$^{-1}$ ($h=0.7$), $\Omega_m=0.3$ for flat $\Lambda$CDM cosmology. Uncertainties are quoted on a 90\% confidence interval unless stated otherwise. Masses are in units of solar mass $\msun$.

\section{Data}
\label{sect:data}
\subsection{Catalogue of dwarf galaxies}
\label{sect:data:galaxy}
Our primary catalogue of galaxies is  MPA-JHU\footnote{\url{https://www.sdss4.org/dr12/spectro/galaxy_mpajhu/}} catalogue of galaxy properties for SDSS DR12 spectroscopic measurements (part of SDSS-IV, \citealt{Alam2015}). The catalogue description is provided in the link mentioned and the techniques used are based on the methods of \citet{Kauffmann2003, Brinchmann2004, Tremonti2004}. This catalogue covers approximately 9300 square degrees of the sky and contains 1472581 sources. 

In the MPA-JHU catalogue, galaxy properties are obtained by fitting galactic spectra with templates of singular stellar populations from the population synthesis code of \citet{Bruzual2003} for different stellar ages and metallicities. The best-fitting model is chosen for a single metallicity and a combination of ten populations of a single age.  The continuum is then subtracted from the data and the residual emission lines are fit with Gaussians.  The result is a set of galaxy parameters, most notable for us are the stellar mass, star formation rate and BPT classification \citep{Baldwin1981}.

The stellar masses are measured with the assumption of the initial mass function of \citet{Kroupa2001} within the SDSS fibre (3 arcsec diameter). The authors also calculate the total stellar mass from the model photometry, i.e. representative of the whole galaxy. We use the median of the total stellar mass estimate as the galaxy's stellar mass. We deem a galaxy as a dwarf candidate if the total stellar mass $\stmass < 10^{9.5} \msun \approx 3\times 10^{9}\msun$. There are 65461 such sources\footnote{39004 of which are on the eastern galactic hemisphere (eROSITA-RU)}. Sources were numbered starting from 0 and this value is used as source ID. Only 'reliable' photometry is used (flag \textsc{RELIABLE=1} from \textsc{galSpecInfo} table). The typical mass uncertainty for this sample is 0.6 dex.  Star formation rate is calculated alike: within the aperture and from the photometry (total star formation rate, following \citealt{Salim2007}). We use the median of the total rate of star formation.

In addition, when possible, we find redshift-independent distance estimation for MPA-JHU galaxies at low redshift ($z<0.01$) from NASA/IPAC Extragalactic Database (NED\footnote{\url{https://ned.ipac.caltech.edu/Library/Distances/}}). Out of a sample of 178 dwarf candidates active in X-rays (see below), 64 have redshift $z<0.01$, and for 29 of those we find the redshift-independent distance estimation in NED. For these galaxies, we use the median of all available measurements for a given galaxy. The aim of this exercise is to correct the redshift stated in the MPA-JHU catalogue for the proper motion of galaxies, which becomes important at $z~<~5\times10^{-3}$ ($1500$ km/s) \citep{Kauffmann2003}. 

\subsection{X-ray data}
\label{sect:data:xray}
The SRG X-ray observatory \citep{Sunyaev2021} is an X-ray mission launched in 2019. In December of that year, it started its all-sky survey which was put to a halt in 2022 after $\approx4.4$ all-sky scans (each scan takes 6 months to complete). The eROSITA X-ray telescope \citep{Predehl2021} operates in the 0.2--9 keV energy range. We use SRG/\ero~catalogue of X-ray sources in the Eastern Galactic hemisphere ($0<l<180\degr$) gathered after over two years of operations (four all-sky scans + 40\% of the fifth all-sky scan). The catalogue was obtained by the X-ray catalogue science working group of the Russian consortium of the eROSITA telescope. Calibration of data, production of sky and exposure maps, and source detection  were carried out using the eSASS software developed by the German SRG/eROSITA consortium \citep{Brunner2022} and the software developed by the Russian SRG/eROSITA consortium. The positional errors and astrometry are calibrated by the catalogue  working group via cross-match with a large sample of optical quasars, resulting in the final and corrected X-ray positions and positional errors. For matching with MPA-JHU dwarf galaxies we used eROSITA X-ray source catalogue in the Eastern Galactic hemisphere constructed in the 0.3--2.3 keV energy range. For further analysis, we filtered the eROSITA catalogue to leave point sources having the 98\% positional uncertainty ($r_{\rm 98}$) better than $20\arcsec$ and with detection likelihood ${\rm DL}>15$, approximately corresponding to $\approx 5\sigma$ significance for Gaussian distribution. We note that in the course of the astrometric correction and error calibration, the 98\% positional errors are capped at the lower limit of $5\arcsec$. Since we are interested in AGN (which would always be a point source), extended objects are not considered in this work.

In the next section, we describe the procedure of matching the eROSITA X-ray catalogue to the sample of dwarf galaxies from MPA-JHU galaxies.

\section{Catalogue of active dwarf galaxies}
\label{sect:catalog}
\subsection{Matching eROSITA X-ray source catalogue with dwarf galaxies.}
\label{sect:catalog:cross-match}

\subsubsection*{Initial cross-match and random-chance associations} 

We start with cross-matching the sample of 65461 SDSS dwarfs (39004 of which are in the eastern galactic hemisphere) with the eROSITA catalogue of sources. The area of MPA-JHU footprint located in the Eastern Galactic hemisphere is 5200 degrees squared. The match radius was chosen to equal 98\% positional uncertainty of the eROSITA source position.  
Sources with negative redshifts in the MPA-JHU catalogue are discarded.  We found 198 
eROSITA-SDSS pairs, made of 198 unique MPA-JHU galaxies and 178 unique eROSITA sources.

To assess the number of random-chance associations between eROSITA and MPA-JHU dwarfs we performed a simple simulation.  eROSITA sources were displaced in a random direction by 
0.1 degree and the cross-match with the original dwarf catalogue was performed. All matches found after random shifts are not real and constitute chance alignments. By repeating the simulation 200 times we estimate that the number of random-chance associations between eROSITA and MPA-JHU dwarf is $33.7\pm5.9$ sources ($\sim17$ per cent, 1$\sigma$ error). The amount of random-chance associations is quite significant due to the large number of eROSITA sources. False matches with X-ray active quasars could in principle be identified with  the help of IR data to exclude quasars from matching pairs. Systematic implementation of this approach is deferred for future work. However, we note that some fraction of false matches will be removed from the sample in the course of the cleaning procedure described below.

\subsubsection*{Removing duplicates}

Among the 198 MPA-JHU objects, there are duplicated pairs of coordinates (when the MPA-JHU have two objects with different IDs, but equal coordinates). We remove one MPA-JHU object from such pairs leaving the one with greater mass\footnote{maximum mass difference in a pair is 6\%}. This removes 11 MPA-JHU sources.

There are still duplicated pairs when one eROSITA source is close to two MPA-JHU entries with different IDs/coordinates. From such pairs, we again remove the one with the smaller mass\footnote{maximum mass difference is a factor of 10-500 in three sources, for 8 sources the figure is of the order of unity.}. During the visual inspection, we confirmed that those 9 cases are, in fact, several close-by SDSS fibre positions measuring the same galaxy (hence the mass difference). This left 178 pairs, with 178 unique X-ray detections and 178 unique SDSS dwarf galaxies. 

Thus those 178 galaxies are our candidate dwarfs with active nuclei.  We significantly increased the purity of the catalogue as per the steps below, but all 178 objects are listed in the tables accompanying this paper.

\subsubsection*{Cleaning the Quasars}

The cleaning continues with the cross-match of the sample with the SDSS DR16 list of quasars \citep{Lyke2020}\footnote{\url{https://www.sdss4.org/dr17/algorithms/qso_catalog/}}. We found 11 distant AGN/quasars within $r_{\rm 98}$ of the eROSITA position and removed them from the active dwarf candidates. Three additional quasars are excluded with the help of Simbad database \citep{Wenger2000}. X-ray emission is much more likely to originate in a quasar rather than (any) galaxy for the eROSITA flux limit.

\subsubsection*{Cleaning of massive galaxies}

In the next step, we cross-correlate the X-ray positions with the whole MPA-JHU catalogue (not only dwarf galaxies) with a 40" radius. We find 19 cases when within $r_{\rm 98}$ we can see a massive galaxy ($\stmass > 10^{9.5} \msun$, independent of the reliability of photometry). We believe that this is an example when a galaxy (usually of a large size) is being measured by several fibres and one of them provides a total stellar mass in the dwarf regime incorrectly (so-called photometric fragmentation, see \citealt{Sartori2015, Birchall2020}). All 19 sources are excluded.

Based on the same arguments we cross-match the sample with NASA-Sloan Atlas of galaxies (NSA \textsc{v1\_0\_1}\footnote{\url{https://live-sdss4org-dr13.pantheonsite.io/manga/manga-target-selection/nsa/}}) catalogue with 40" radius, and remove galaxies which have a massive $\stmass > 10^{9.5+0.3} \msun$ neighbour within $r_{\rm 98}$. Note that we increase the dwarf mass limit by 0.3 dex (factor of 2) to conservatively accommodate the mass measurement uncertainty. 45 objects are excluded in this manner (all except a few sources are photometric fragmentation examples). Some of them also have a massive neighbour from SDSS (see step above).

\subsubsection*{Visual inspection}

\begin{figure*}
    \centering
    \includegraphics[width=\hsize]{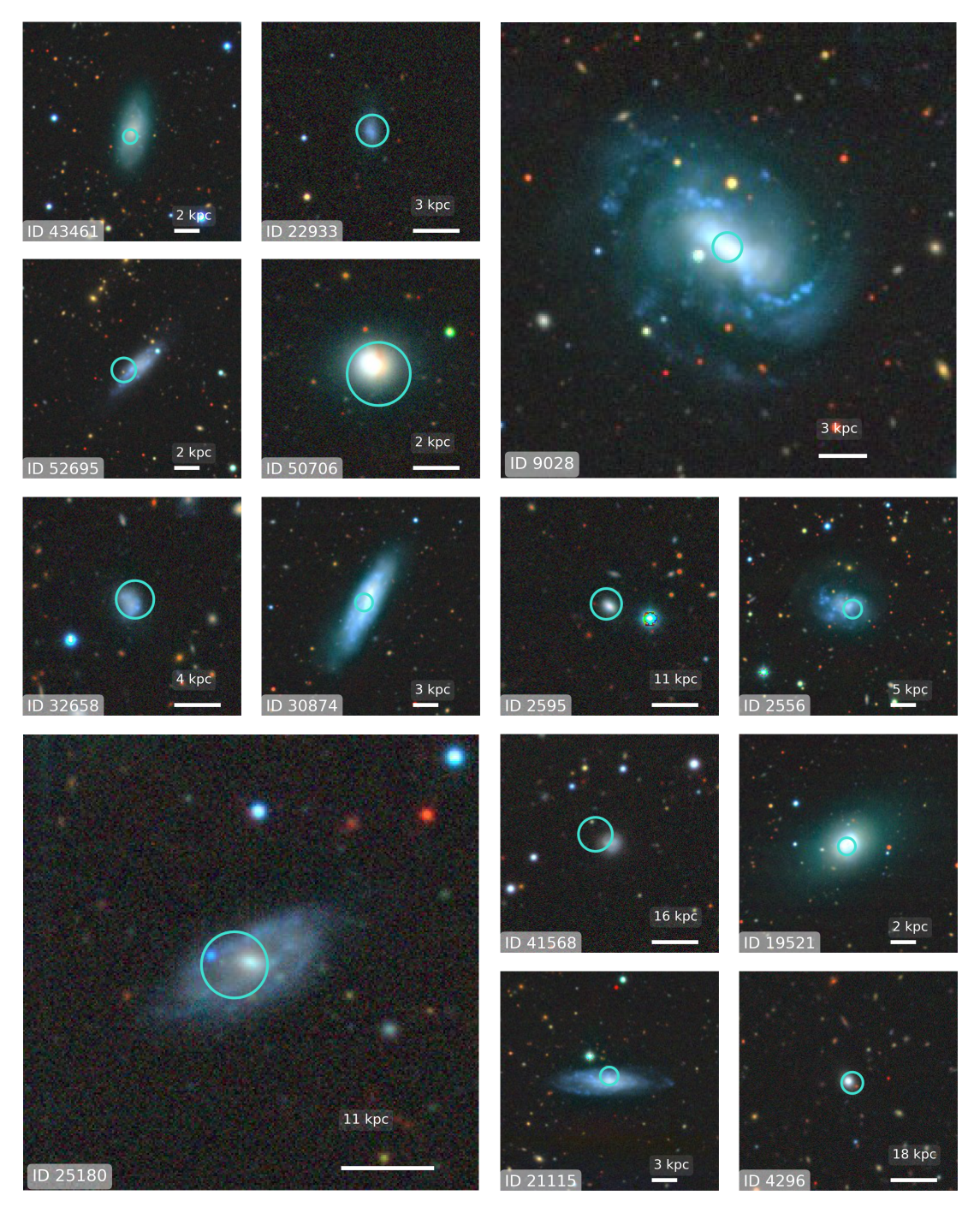}
    \caption[Examples of dwarf galaxies with X-ray activity detected by eROSITA.]{Examples of galaxies which passed the selection criteria. Each image is a DESI LIS cutout centred on the dwarf's position. eROSITA X-ray source position and positional uncertainty ($r_{98}$) shown as cyan circle. Each galaxy has an ID on the bottom left and the scale bar on the bottom right (20" in length).}
    \label{fig:stamps}
\end{figure*}

The final step in the cleanup is the visual inspection of each source and its surroundings using DESI LIS\footnote{\citealt{Dey2019}, \url{https://www.legacysurvey.org}} cutouts. In this step, we excluded 16 sources. 15 cases were examples of photometric fragmentation or HII regions in large  (12 objects) or dwarf (3 objects) galaxies outside of $r_{\rm 98}$, and one case of a stellar object (Simbad). This leaves us with 99 remaining  and 79 excluded sources. Examples of AGN dwarf candidates are presented in Fig. \ref{fig:stamps}, and examples of excluded sources due to different reasons are presented in Fig. \ref{fig:stamps-bad}. In the attached tables, a comment is made on the reason for the exclusion of removed sources.

Even though we excluded almost half of eROSITA matches, the catalogue may not be 100\% pure and extreme objects (e.g. in mass or luminosity) should be treated with caution. For example, we note that 9 out of 10  galaxies with extremely low stellar mass ($M_*<10^7\msun$) have counterparts in the NSA catalogue. Out of 9 such galaxies, 7 have a significantly more massive dwarf counterpart in NSA (e.g. for ID 28027 the mass from MPA-JHU is $\sim10^6$ and from NSA $\sim6\times10^8\msun$), therefore properties of such sources should be used with care.  For the overall consistency, we use only MPA-JHU masses in this work.

In what follows we will refer to the sources which passed the cleaning criteria unless stated otherwise.

\subsection{X-ray variability and spectral analysis}
\label{sect:catalog:xray}
For each source we extract data during each of four (five) eROSITA all-sky surveys (eRASS1...5) and the co-added data of all surveys. 61 sources have data from the fifth sky survey. For each scan (or added data) the source extraction region is a circle with a $60\arcsec$ radius centred on the position of an X-ray source from the added data. The background was estimated from an annulus with inner and outer radii of $120\arcsec$ and $300\arcsec$. From both regions, we exclude a $40\arcsec$ circle around any source detected nearby (with the detection likelihood ${\rm DL}>10$). The spectra were rebinned to have at least three source counts per channel.  The fitting is done in \textsc{Xspec} \citep{Arnaud1996} with Cash statistics \citep{Cash1979} in the 0.3--8 keV energy range. Spectral models are not red-shifted due to the relatively small distances involved.

\subsubsection{Spectral analysis and X-ray flux calculation}

In order to characterise the range of spectral properties and to compute X-ray flux, we fit the X-ray spectra of all candidates with a simple power law model absorbed with the neutral hydrogen in our Galaxy. The hydrogen column density $N_{\rm H}$ is fixed for a given source on the value from the NH4PI map at the source's position \citep{HI4PICollaboration2016}. Sources in our sample have a fairly low number of counts (the median of net source counts is $\sim20$) and correspondingly large uncertainties in their spectral parameters. To avoid additional errors due to uncertainties in spectral parameters, for the flux calculation we used several  models of fixed spectral shape chosen depending on the power law fit results, as described below.

\begin{figure}
    \includegraphics[width=0.5\textwidth]{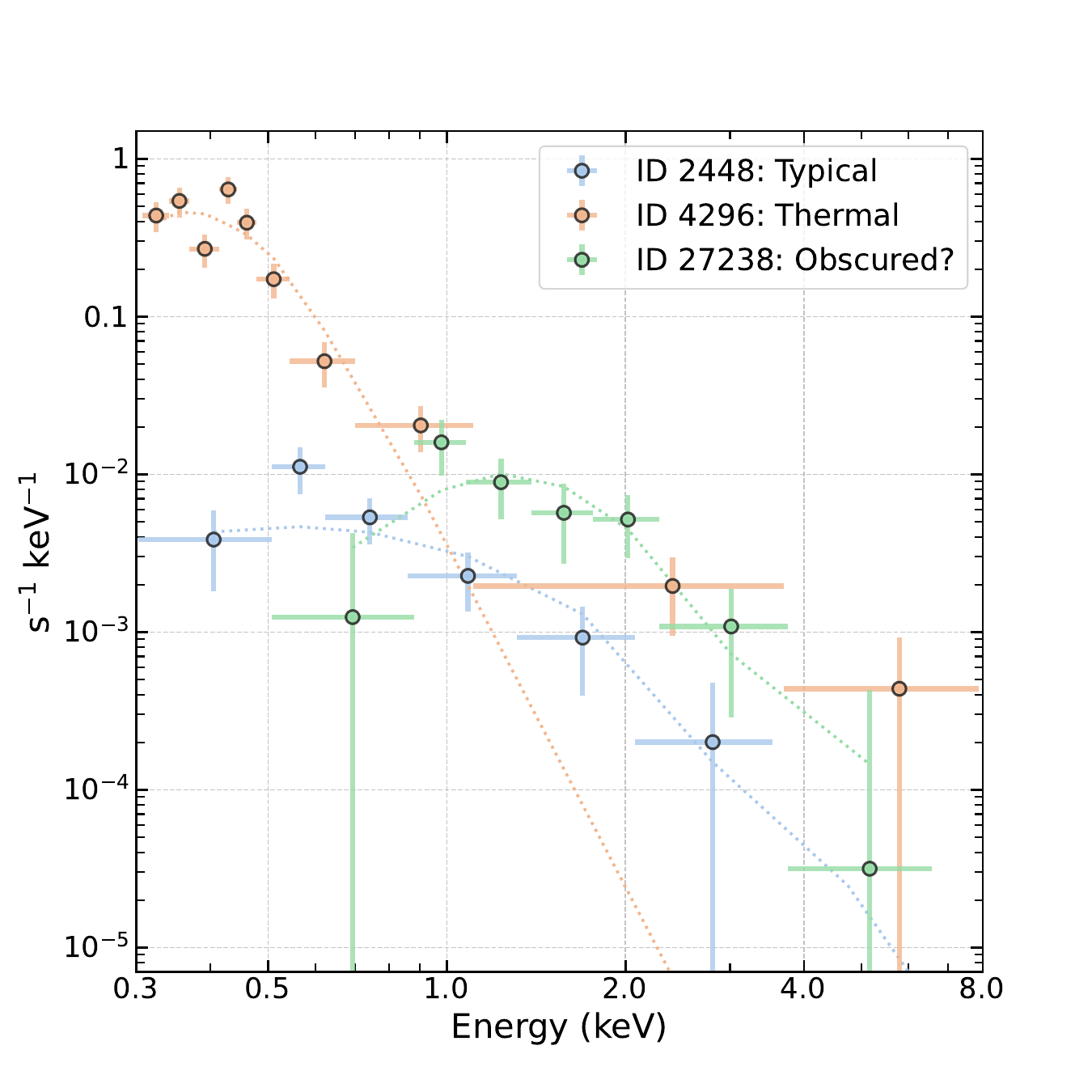}
    \caption[Representative X-ray spectra of dwarf galaxies as seen by eROSITA.]{Representative X-ray spectra of dwarf galaxies as seen by eROSITA. Three sources are shown - an example of a 'typical' object (modelled with a power law, coloured blue), a thermal spectral shape (modelled with a black body model, orange), and a seemingly absorbed spectrum (modelled with a power law with additional absorption, green). Spectra were rebinned for plotting purposes.  The model used to calculate fluxes is shown with dotted lines.}
    \label{fig:spectra}
\end{figure}

86 out of 99 sources have photon indices consistent (within 90\% confidence interval) with the canonical AGN values in the range $1.5\lesssim\Gamma\lesssim2.5$ \citep{Ge2022, Liu2022}. For those sources, we used a fiducial spectral model of a power law with fixed photon index $\Gamma = 1.9$ for flux calculation.  Three sources\footnote{ID 10487, 33862 and 4296} are statistically significantly softer than this range, we used a black body model \textsc{bbodyrad} with frozen best-fit temperature and with fixed Galactic absorption for them\footnote{In X-ray spectra of ID 10487 and 4296 may have an additional power-law component as described in sect. \ref{sect:indiv:soft_agn} and \ref{sect:indiv:tde} respectively.}. Spectra of ten sources are harder than the typical AGN spectrum with the 90\% confidence interval for the photon index $\Gamma\lesssim1.5$. Spectra of 8 of them\footnote{ID 52047, 51004, 50909, 45689, 27238, 5545, 20944 and 11124} can be adequately described by the model of a mildly absorbed AGN with $\Gamma=1.9$ and $N_{\rm H}=5\times10^{21}$ cm$^{-2}$ and we used this model for their flux calculation. Spectra of two remaining sources\footnote{ID 15677 and 49241} are much harder and could not be  fit with the mildly-absorbed $\Gamma=1.9$ power law. For the flux calculation of these sources we used a power law with  $\Gamma=1.0$ and Galactic absorption.
Examples of representative spectral shapes are shown in Fig. \ref{fig:spectra} for three relatively bright sources.

With the chosen fiducial models, the source's flux and luminosity were calculated in the 0.3--2, 0.5--2, and 2--8 keV energy ranges\footnote{For ID 15677 and 25312 luminosity errors are quoted on 68\% uncertainty level, since 90\% errors formally give only upper limits.}.

\subsubsection{Variability}

We analysed the light curve of each source by calculating flux in the 0.3--8 keV energy range for each scan. Flux (or its upper limit) was calculated using spectral models described above with the shape parameters fixed (e.g. $\Gamma$ or black body temperature).  4 sources have only upper limits in all scans.
Dwarfs were ranged by the ratio of the maximum measured flux to the minimal flux/upper limit ignoring error bars. We call this ratio $R$ and use it as an approximate variability indicator.

\begin{figure}
    \includegraphics[width=0.5\textwidth]{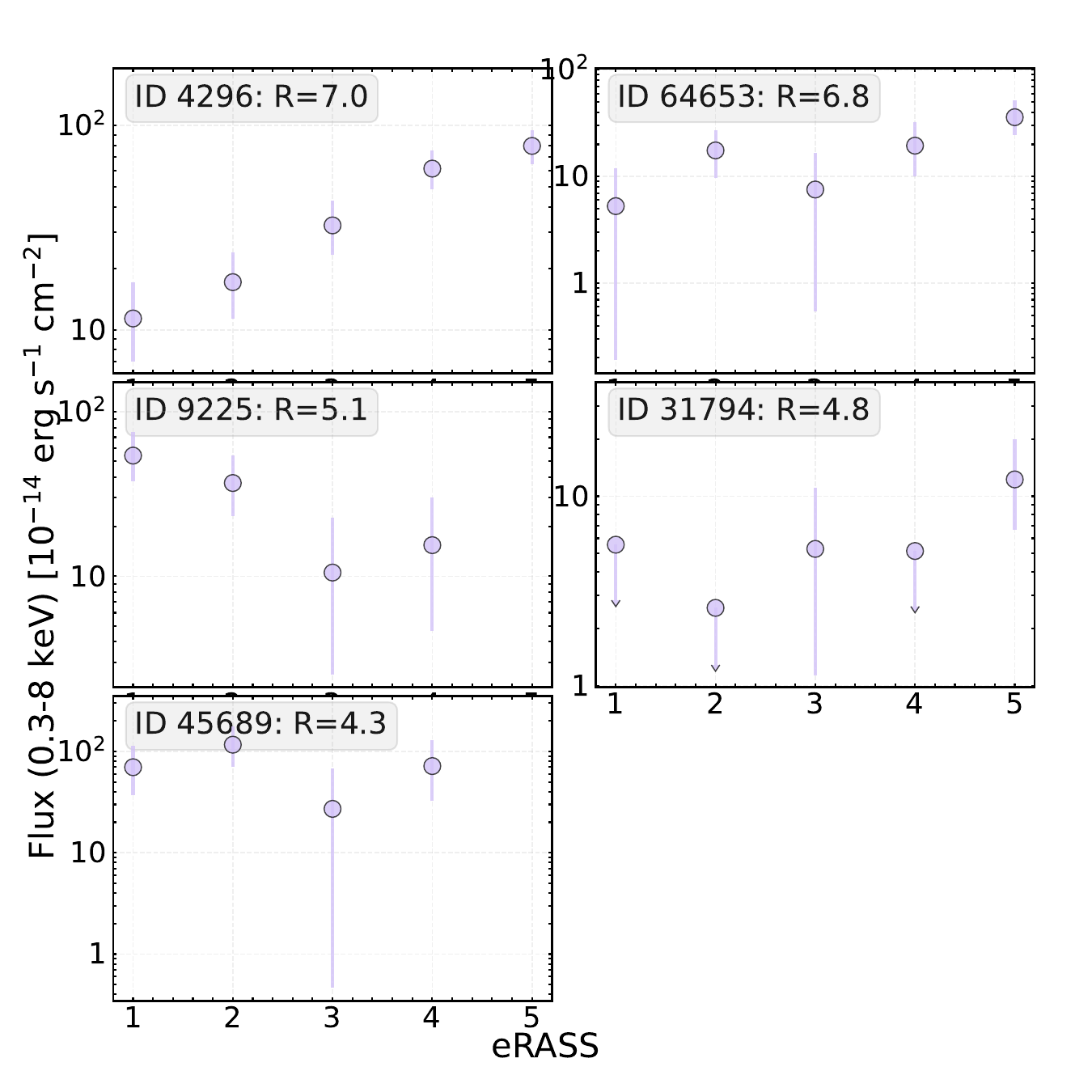}
    \caption[X-ray light curves of five dwarf galaxies.]{X-ray light curves of five dwarf galaxies with highest values of R. Each point shows the  average flux registered in one all-sky survey. }
    \label{fig:variability}
\end{figure}

Out of 99 dwarf candidates, only 3 have  $R>5$, i.e. suspected significantly varying sources.  Examples of light curves are shown in Fig. \ref{fig:variability}, where the top 5 variable sources are shown.

For  more thorough techniques to search for variability suited for eROSITA (including upper limits and proper treatment of $R$ uncertainties) see, e.g. \cite{Buchner2022, Medvedev2022}. Using those, however, is deferred for future work.

\subsection{Contribution of X-ray binaries and hot gas}
\label{sect:catalog:xrb}

X-ray emission in galaxies can come from a multitude of sources, but the primary contaminant for our study would be the X-ray emission from X-ray binaries and hot gas in the interstellar medium (ISM) \citep{Mineo2012b, Gilfanov2022}. The binary population consists of two types of sources, Low- and High-mass X-ray binaries (LMXB and HMXB), in which a black hole or a neutron star is paired with a normal star. LMXBs trace the old stellar population, whilst HMXB stalk young massive stars
\citep{Grimm2003, Lehmer2010, Mineo2012, Mineo2012b, Gilfanov2022}. 

The aggregated emission of both binaries populations and hot gas is tightly correlated with the main galaxy properties -- stellar mass and star formation rate. Namely, the number and total luminosity of LMXB systems depend on the stellar mass, whereas that of HMXB depends  on the star formation rate (SFR). The luminosity of hot gas  is also proportional to the SFR. To predict the contributions of binaries populations for our dwarf candidates we use scaling relations from \citet{Lehmer2010} and \citet{Gilfanov2022}\footnote{namely, we use formula (5) from \citealt{Gilfanov2022} with luminosities in the 0.5--8 keV energy range and stellar mass/SFR from the MPA-JHU catalogue. $L_{\rm X, XRB}=\beta\times\text{SFR} + \alpha\times M_*$, $\log\alpha=29.25$, $\log\beta=39.71$ \citep{Lehmer2010}}. For hot gas, we use the relation from \citet{Mineo2012b}\footnote{$L_{\rm gas}=8.3\times10^{38}\times$SFR, with luminosities in the 0.5--2 keV range.}. 

Out of 99 active dwarf candidates, 82 have the estimated XRB contribution ($L_{\rm X,XRB}$) far below the observed luminosity ($L_{\rm X,obs}$) in the 0.5--8 keV band, $\lxlxrbrat>3$. In all of those 82 sources the contribution of hot gas is also negligible compared to the observed luminosity in the 0.5--2 keV energy band. We conclude that the majority of our X-ray active dwarf galaxies are  indeed powered by  accretion onto a massive black hole. It is interesting to note that dwarf galaxies   with $\lxlxrbrat<3$  do not show prominent X-ray variability, with the largest $R=2.8$ for ID 39612.

\begin{figure}
    \includegraphics[width=0.5\textwidth]{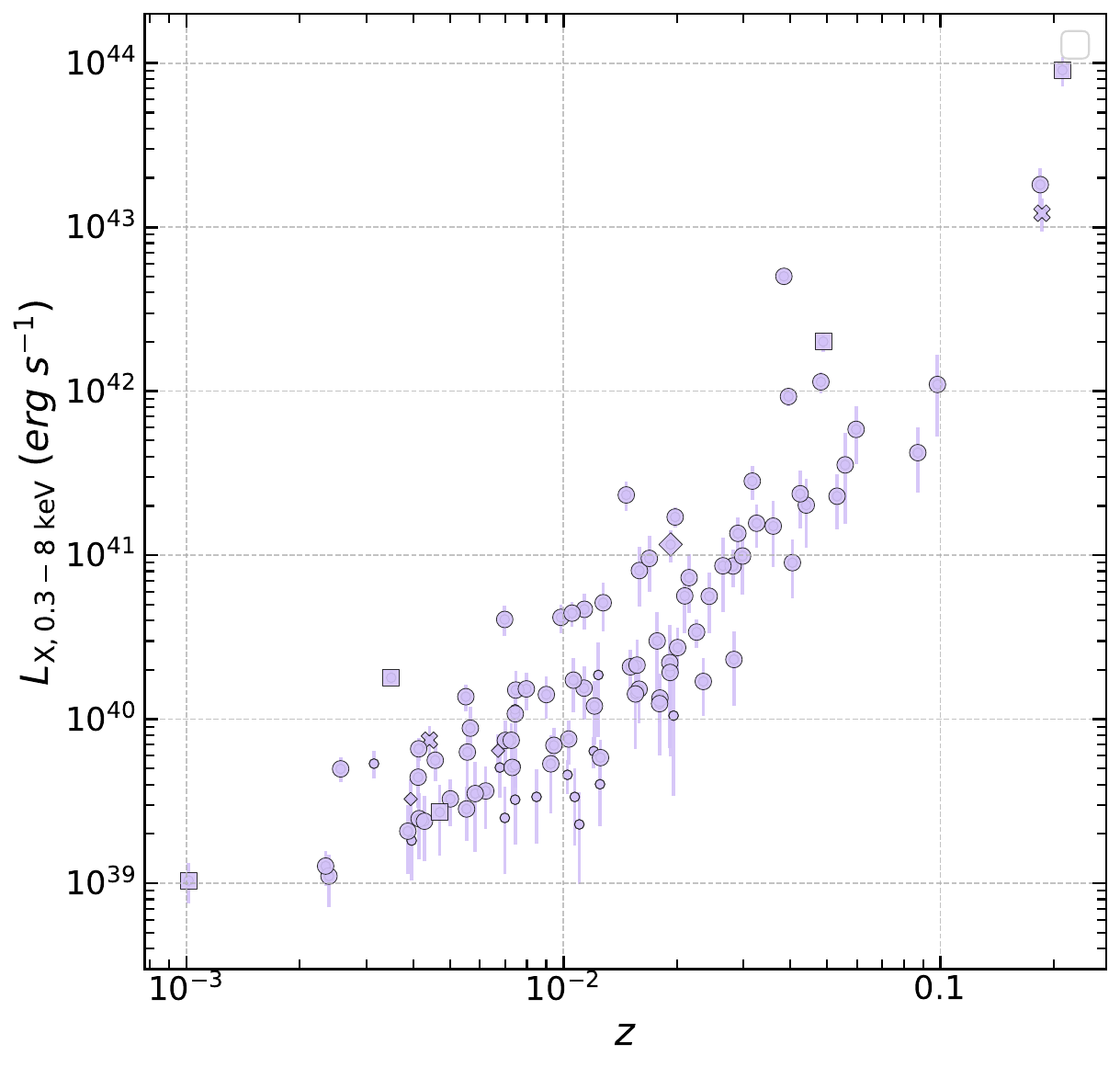}
    \caption[Dwarf galaxies redshift - X-ray luminosity plane.]{Redshift v. X-ray Luminosity plane for 99 dwarf galaxies detected by eROSITA. Larger-sized markers designate objects with $\lxlxrbrat>3$. Marker encodes dwarf's BPT classification from MPA-JHU (circle -- star-forming, rhombus -- composite, square -- AGN, unclassified -- X). }
    \label{fig:Lx_z}
\end{figure}

\section{Results \& Discussion}
\label{sect:results}

\subsection{The catalogue of X-ray active dwarf galaxies and candidates}

In appendix \ref{appendix:catalog} we describe the catalogue  of 99 X-ray active  dwarf galaxies detected by eROSITA, and present a subset of its  columns in table \ref{tab:dwarfs} for dwarfs with $\lxlxrbrat>3$.
The list of 76 eROSITA -- MPA-JHU matches which were not included in the final catalogue is presented in appendix \ref{appendix:excluded} along with the reason for rejection for each case. DESI LIS images are available for all objects from the files attached to the paper.

Fig. \ref{fig:Lx_z} shows the redshift -- X-ray luminosity  plane, with markers encoding the BPT classification of an object (see below) and the XRB contribution. The minimum luminosity is $\sim10^{39}$ \ergps,  the maximum is $\sim10^{44} $ \ergps, median is $1.5\times10^{41}$ \ergps. All sources except a few are on redshift $z<0.1$, with the median of $z=0.01$. The correlation between redshift and luminosity is expected for a flux-limited sample of eROSITA, but AGNs (according to the BPT classification) tend to be more luminous than SF galaxies.

\subsection{Objects previously identified as AGN candidates}
\label{sect:results:known_dwarfs}

Before turning to the discussion of the results, we cross-correlated (5" match radius) our dwarf sample with some readily available active dwarf candidates catalogues mentioned in the introduction. First,  the sample of 61 X-ray-selected active dwarfs from  3XMM-DR7 \citep{Birchall2020} to find only 5 matches (IDs 11124, 20701, 34500, 49241, 53431).  The study of \citeauthor{Birchall2020}, to our knowledge, used the largest publicly available X-ray survey with the area of $\sim1000$ deg$^2$ -- the closest one gets to eROSITA's power. We found four dwarf with AGN (ID 9225, 20701, 28027 and 49241) from 19 candidates of  \citet{Lemons2015}. Four optically-selected active dwarfs from \citet{Reines2013} may be found in our catalogue  (ID 53431, 10487, 44628, 49241). Optical + mid-IR selection by \citet{Sartori2015} matched with the following 6 sources - ID 44628, 25337, 10487, 30754, 9028, 34500, 4296. Infrared selection of \citet{Marleau2017} coincided with our ID 10487, 44628, 20701, 20867, 11124, 49241. Finally, we searched for X-ray emission from 305 IMBH candidates of  \citealt{Chilingarian2018} to find three matches (ID 10487, 49241, 53431). From \citet{Greene2007} we re-discovered only ID 10487.

Overall, 14 unique eROSITA-detected dwarfs were reported as (candidate) AGNs in low-mass galaxies before. Leaving our sample with $\sim85$ per cent new candidates discovered for the first time.

ID 6328, 19521, 20867, 33862 and 50706 are found in a sample of  dwarf galaxies targeted for radio observation in \citet{Reines2020}. They used high-resolution radio observations to establish that a large number of massive black holes in dwarfs are off-centred. Only ID 19521  and 20867 have compact radio sources near the galactic centre (offsets 0.3" and 0.6" whereas positional uncertainty is $<0.1"$, indicating offset from the centre). However both galaxies have radio properties consistent with star formation processes (see fig. 11 in \citeauthor{Reines2020}, our galaxies are their ID 49 and 62 respectively).

\subsection{Sample properties}
\label{sect:results:optical}
\subsubsection{Stellar mass, SFR and the BPT-diagram}
\label{sect:results:optical:bpt}
\begin{figure}
    \includegraphics[width=0.5\textwidth]{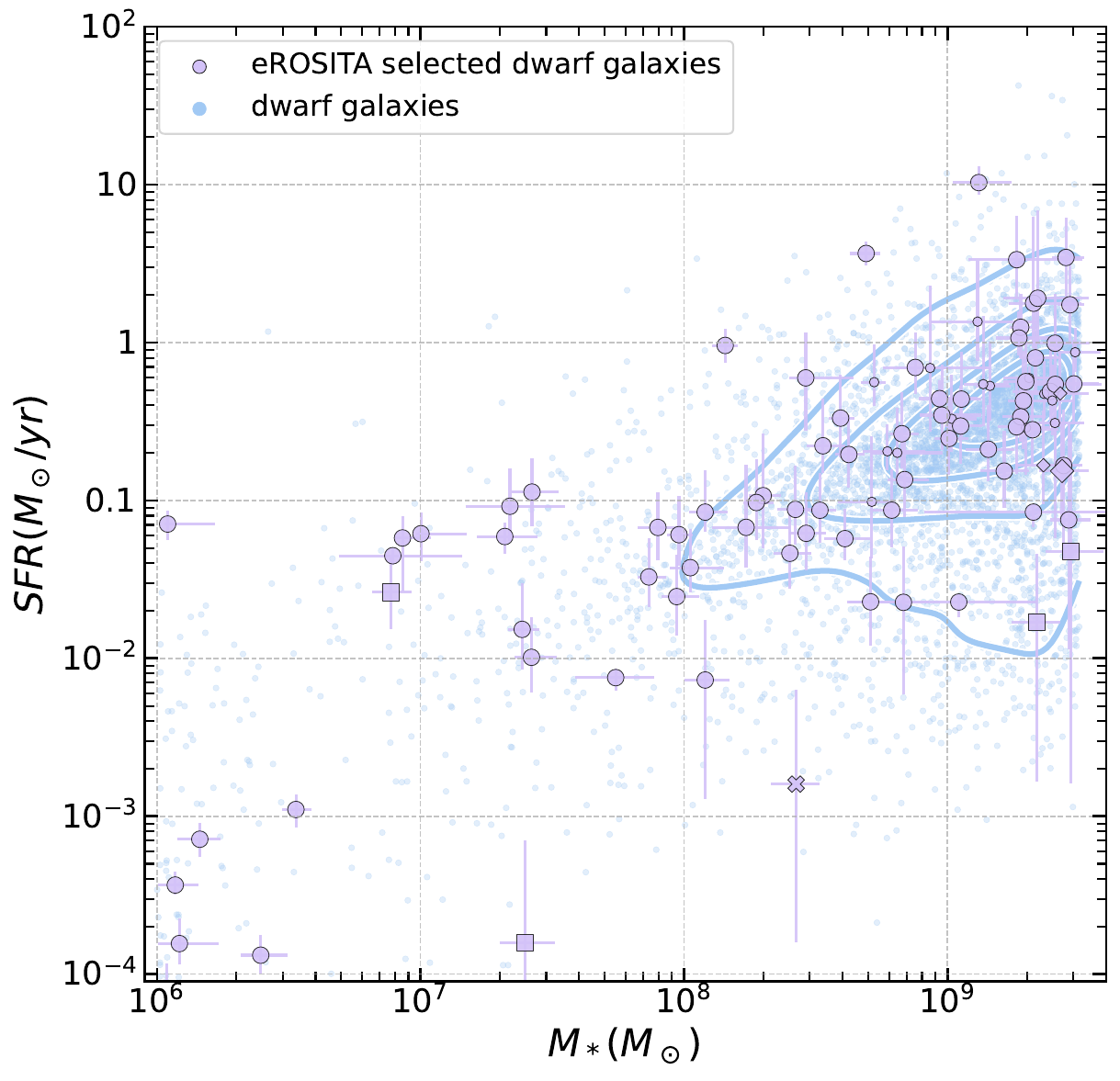}
    \caption[Dwarf galaxies stellar mass - star formation rate plane.]{Position of 99 eROSITA-detected dwarf galaxies in the Stellar Mass v. Star Formation Rate plane (purple circles). Larger-sized markers designate objects with $\lxlxrbrat>3$. Marker encodes dwarf's BPT classification from MPA-JHU (circle -- star-forming, rhombus -- composite, square -- AGN, unclassified -- X). In addition, 5000 dwarf galaxies from MPA-JHU selected at random are shown as blue circles. Contours show iso-proportion density (inner -- 20, 40, 60, 80\%  -- outer) for 20k random dwarfs. It is apparent that eROSITA-selected dwarfs have more objects of lower mass than that expected from the contours of the parent population.}
    \label{fig:M_sfr}
\end{figure}

We start describing the catalogue of 99 selected dwarfs with the description of host properties. In the figures, data from MPA-JHU is shown with $1\sigma$ uncertainties.
In Figure \ref{fig:M_sfr} we show the distribution of objects in $M_*$-SFR plane. For eROSITA dwarfs we show their BPT classification as different markers (see below). 
$\sim80$ per cent of the sample have stellar mass $M_*>10^{8} \msun$. The lowest mass is $10^6 \msun$. Star formation rate spans from $10^{-4}$ to $10$ $\msun$ yr$^{-1}$. Medians are $7\times10^8 \msun$ and $0.15$ $\msun$ yr$^{-1}$ respectively. A large sample of dwarfs from MPA-JHU is also shown in the figure. KS-test for the distribution of masses show that the masses of eROSITA-selected objects and dwarf galaxies\footnote{eastern galactic hemisphere} are not  drawn from the same distribution (p-value $\sim10^{-4}$) - eROSITA objects have a slightly heavier low-mass tail. The same test for SFR values shows the p-value of $0.02$, marginal consistency with the same distribution. 

\begin{figure}
    \includegraphics[width=0.5\textwidth]{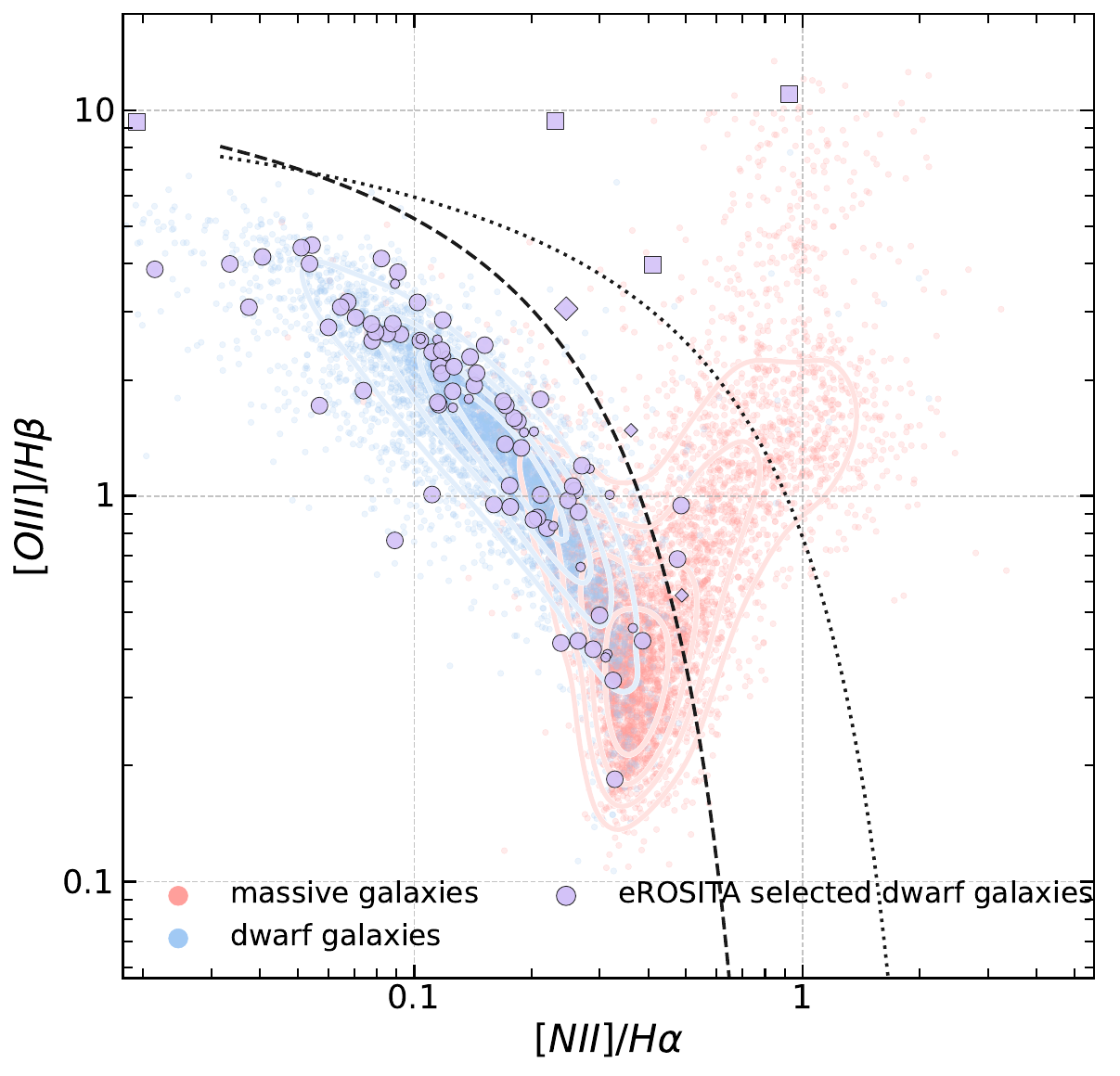}
    \caption[[Dwarf galaxies BPT diagram.]{Position of eROSITA-selected dwarf galaxies in the BPT diagram. Colours and symbols are the same as in Fig. \ref{fig:M_sfr}, but with added 5000 random massive (non-dwarf) galaxies from MPA-JHU (red). SF-Composite-AGN separation is shown as black dashed and dotted lines.}
    \label{fig:bpt}
\end{figure}

In Fig. \ref{fig:bpt} we show the BPT-diagram \citep{Baldwin1981}. It shows the ratio of two pairs of optical emission lines - $\rm \frac{O_{III, 5007}}{H_{\beta}}$ and $\rm \frac{N_{II, 6583}}{H_{\alpha}}$. The diagram is a diagnostics tool for the source of ionising radiation and is good in separating optical AGN and galaxies with active star formation. We plot the eROSITA sample, and samples of randomly selected dwarfs and massive galaxies from MPA-JHU, with all samples filtered to have the necessary line fluxes signal-to-noise larger than three (92 X-ray active dwarfs pass this criterion). For eROSITA sources their BPT classification from MPA-JHU (methodology of \citealt{Brinchmann2004}) is shown. Separation lines from \citealt{Kewley2001, Kauffmann2003b} between star-forming (SF), composite and active galaxies are shown. eROSITA sample has 89 SF galaxies, 5 AGN, 3 composite and 2 unclassified sources. Both eROSITA-selected and MPA-JHU dwarfs are dominated by SF galaxies, whereas massive objects are more dispersed in both SF and AGN branches. This plot shows that X-ray emission is able to pick up low-luminosity AGNs deep in the SF region of the BPT diagram, as explained in the introduction and consistent with previous X-ray studies (e.g. \citealt{Birchall2020}). Additional diagnostics can be performed based on the ratio of X-ray flux to the optical emission flux. For example the ratio of $\rm O_{ III, 5007}$ line to the hard X-ray luminosities is used to discern AGNs from TDE in eROSITA survey \citep{Khorunzhev2022}, or the ratio of the flux at $2500$ angstrom to the flux at 2 keV \citep{Birchall2020}; but we do not attempt such diagnosis here.

\subsubsection{X-ray luminosity and Eddington ratio}
\label{sect:results:optical:eddington}

\begin{figure}
    \includegraphics[width=0.5\textwidth]{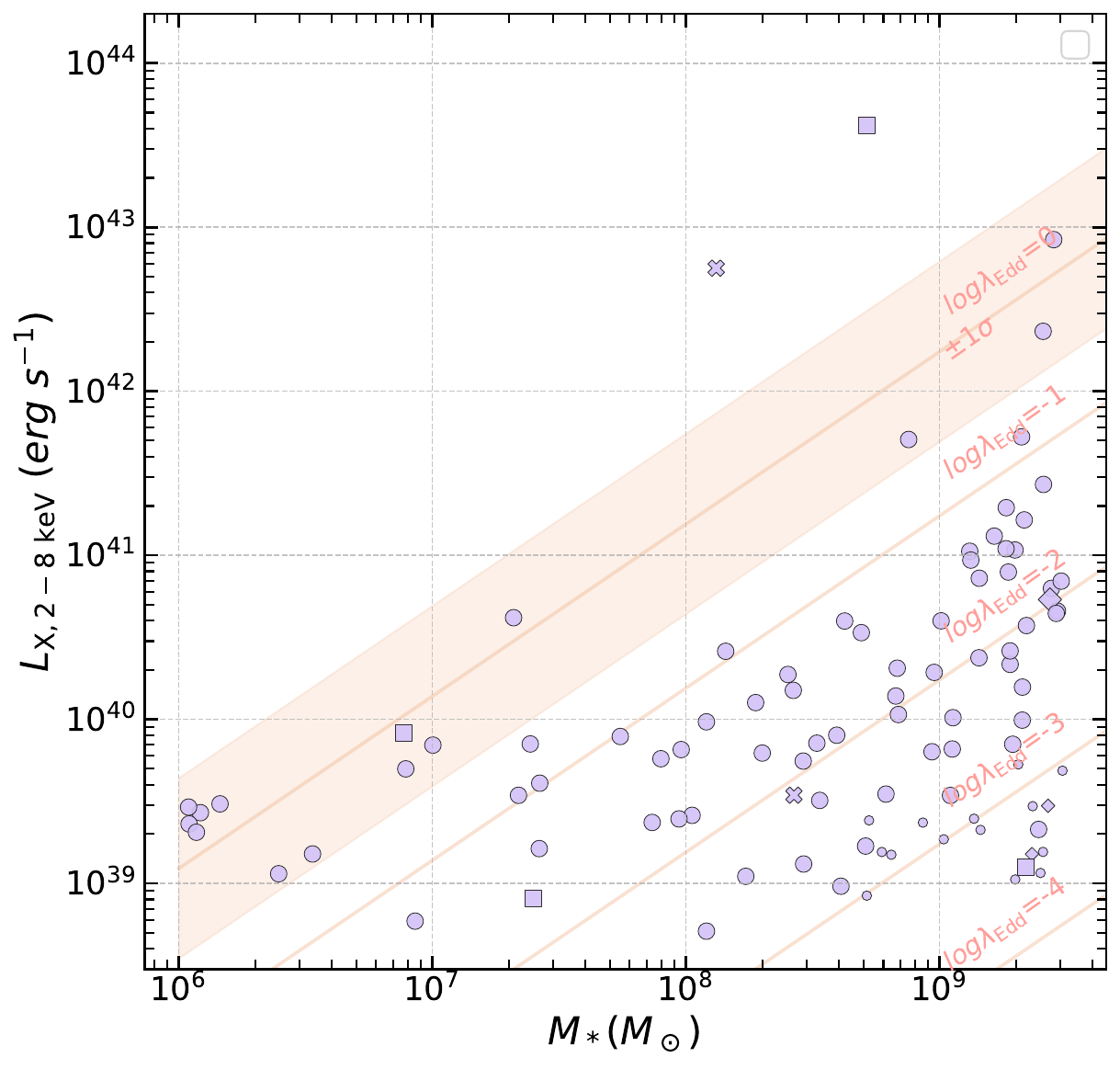}
    \caption[Dwarf galaxies stellar mass - X-ray luminosity plane.]{Mass v. X-ray Luminosity plane. Lines of constant Eddington ratio are shown with orange lines. For the $\lambda_{\rm Edd}=1$ the expected scatter in $M_{\rm BH}-M_*$ relation is shown. Points above $\lambda_{\rm Edd}=0.1$ should be interpreted with extreme care (see text). }
    \label{fig:Lx_M}
\end{figure}

Figure \ref{fig:Lx_M} shows the hard X-ray luminosity (2--8 keV) vs. host stellar mass. We plot lines of constant Eddington ratio $\lambda_{\rm Edd}$ \citep[see][]{Aird2012, Birchall2020, Birchall2022, Zou2023} calculated as explained below.

We define
$$
\lambda_{\rm Edd} = \frac{K\times L_{\rm 2-10~keV}}{1.3\times10^{38} M_{\rm BH}(M_*)}
$$

where $K$ is the bolometric correction (translates hard X-ray luminosity into bolometric luminosity), $M_{\rm BH}(M_*)$ is a mass of a central black hole expected for a given stellar mass. The coefficient in the denominator is the Eddington luminosity for unit mass.
For an X-ray spectrum with photon index $\Gamma=1.9$, the flux in the 2--10 keV band is 17\% higher than in 2--8 keV. Given uncertainties in bolometric corrections and black hole masses, we ignore this difference and use measured 2--8 keV luminosities instead. No k-correction is made because redshifts are small.

AGN hard X-ray bolometric corrections are calibrated with respect to the bolometric luminosity, Eddington ratio and black hole mass in \citet{Duras2020}. As we work in the low-mass regime, we use Eq. 7 from \citeauthor{Duras2020} to calculate the bolometric correction for low BH masses. From their Fig. 9 and Eq. 7 one may estimate the $K\approx 16.7$ with the scatter of around $0.35$ dex. Similar X-ray studies also use $K$ in the range of 20-25 \citep{Birchall2020, Latimer2021, Zou2023}. 

For black hole masses, we use scaling relations of \citealt{Reines2015} (their eq. 4, see also \citealt{Suh2020}) calibrated with AGN and galaxies in the local universe, including dwarf galaxies:
$\log(M_{\rm BH}/\msun) = 7.45 + 1.05\log\frac{M_*}{10^{11}~\msun} \pm 0.55$. We note that according to such a relation, objects with $M_*~<~3\times10^9 \msun$ are expected to host a BH of intermediate mass ($M_{\rm BH}~<~10^6 \msun$).

Lines of constant $\lambda_{\rm Edd}$ are shown in Fig. \ref{fig:Lx_M}.  The majority of our sources have $\lambda_{\rm Edd}=10^{-3}...10^{-1}$, with around $\sim10-15$ sources seemingly accreting at the Eddington level or above, median $\lambda_{\rm Edd}$ is $\sim0.01$. The median level is lower than that found in deep surveys such as Chandra \citep{Mezcua2018} and XMM \citep{Zou2023} with median values around $\sim0.2-0.6$. This may be explained by the selection effects - in the distant Universe one has trouble detecting low-$\lambda_{\rm Edd}$ source due to them being under the flux limit. On the other hand, our median is higher than that of \citet{Birchall2020} with the median of $\sim0.001$. The latter may be explained by the $M_{\rm BH}(M_*)$ relation adopted by \citeauthor{Birchall2020} which predicts BH mass almost order of magnitude larger than the relation we adopt, our results are broadly consistent.

There are important caveats in interpreting the results. First, in the high $\lambda_{\rm Edd}$ regime the bolometric correction may change dramatically \citep{Duras2020}. Not only that, the spectral shape may be different from the expected power-law. In our sample two very bright sources ($L_{\rm X, 0.3-2.0 \text{ keV}}\sim10^{42}$) have thermal spectrum, therefore having negligible luminosity in the 2--8 keV regime. These two sources are basically missing in the calculations of $\lambda_{\rm Edd}$ presented here, but still even without bolometric correction they appear to have $\lambda_{\rm Edd}\sim0.1$. One should also bear in mind that the correlation between the stellar and BH masses is highly uncertain in the low-mass regime \citep{Suh2020, Mezcua2023}.  Very low-massive galaxies ($M_*<10^7\msun$), and sometimes normal dwarf galaxies may be prone to mass measurement errors (see sect. \ref{sect:catalog:cross-match}).

\subsection{AGN fraction and luminosity distribution}
\label{sect:results:optical:fraction}

In order to compute the AGN fraction we need to take into account several incompleteness factors. The first one is the  sensitivity of the  eROSITA all-sky survey (varying across the sky) and the flux limit of its source catalogue. This factor is fully accountable for and will be corrected as described below. The second incompleteness factor is related to the incompleteness of the MPA-JHU SDSS catalogue at the faint (low mass, large redshift) end.  We will attempt below several approaches to approximately correct for this effect and demonstrate that they all give consistent results.

Another incompleteness factor is related to the cleaning of the initial list of eROSITA -- MPA-JHU matches.  
Our catalogue of X-ray active dwarf galaxies is constructed with the goal of being pure but not necessarily  complete.
Hence, we are likely to determine only the lower limit on the AGN fraction in dwarf galaxies. This factor is difficult to accurately correct. We note, however, that  96 sources out of 178 initial matches were excluded from the luminosity function calculation (after the cleaning procedure and the requirement of $\lxlxrbrat>3$). Therefore, the maximal uncertainty which can be introduced in this step is less than a factor of $\sim 2$; this should be taken into account when interpreting our results. Finally, an important question is how well  the MPA-JHU catalogue  represents the overall population of dwarf galaxies in the local Universe. This is discussed in more detail at the end of this section.

For calculating AGN fraction and luminosity distribution, only dwarfs with $\lxlxrbrat>3$ are used (82 objects). From MPA-JHU galaxies we removed galaxies duplicated in \textsc{specObjID} or coordinates, and sources with non-reliable photometry.  We limit the stellar mass to the range $8<\log{M_*}<9.5$ - the completeness limit is not well-defined for lower mass galaxies due to poor statistics.


To correct for the incompleteness of the X-ray detections we used a method similar to the approach used in the incompleteness correction of the $\log N - \log S$ flux distribution of X-ray sources \citep{Shtykovskiy2005, Voss2009}. We use the eROSITA sensitivity map to compute the sensitivity curve $A(L_{\rm X})=N(L_{\rm X, upper}<L_{\rm X})$ defined as the number of dwarf galaxies in the given mass-redshift range in the MPA-JHU catalogue where the eROSITA sensitivity is better than a given luminosity $L_X$. Next, for every eROSITA-detected dwarf galaxy  with luminosity $L_{\rm X, k}$  we calculate the weight $w_{\rm k} = 1/A(L_{\rm X, k})$.

The binned estimate of XLF in a bin of luminosity would be then $$\phi = \frac{\dd N}{\dd\log{L_{\rm X}}} = \frac{1}{\Delta\log{L_{\rm X}}} \sum_{\rm k\in \text{bin}}{w_{\rm k}}$$ and its error $\Delta\phi= \frac{1}{\Delta\log{L_{\rm X}}}\sqrt{\sum{w_{\rm k}^2}}$. For the cumulative XLF, $$\Phi(L_{\rm X})=\sum_{\rm k: L_{\rm X, k}>L_{\rm X}}{w_{\rm k}}$$
When no sources are detected in a bin/above a given luminosity, an upper limit of 95\% is placed which for Poissonian distribution equals to three sources.

\begin{figure}
    \includegraphics[width=0.5\textwidth]{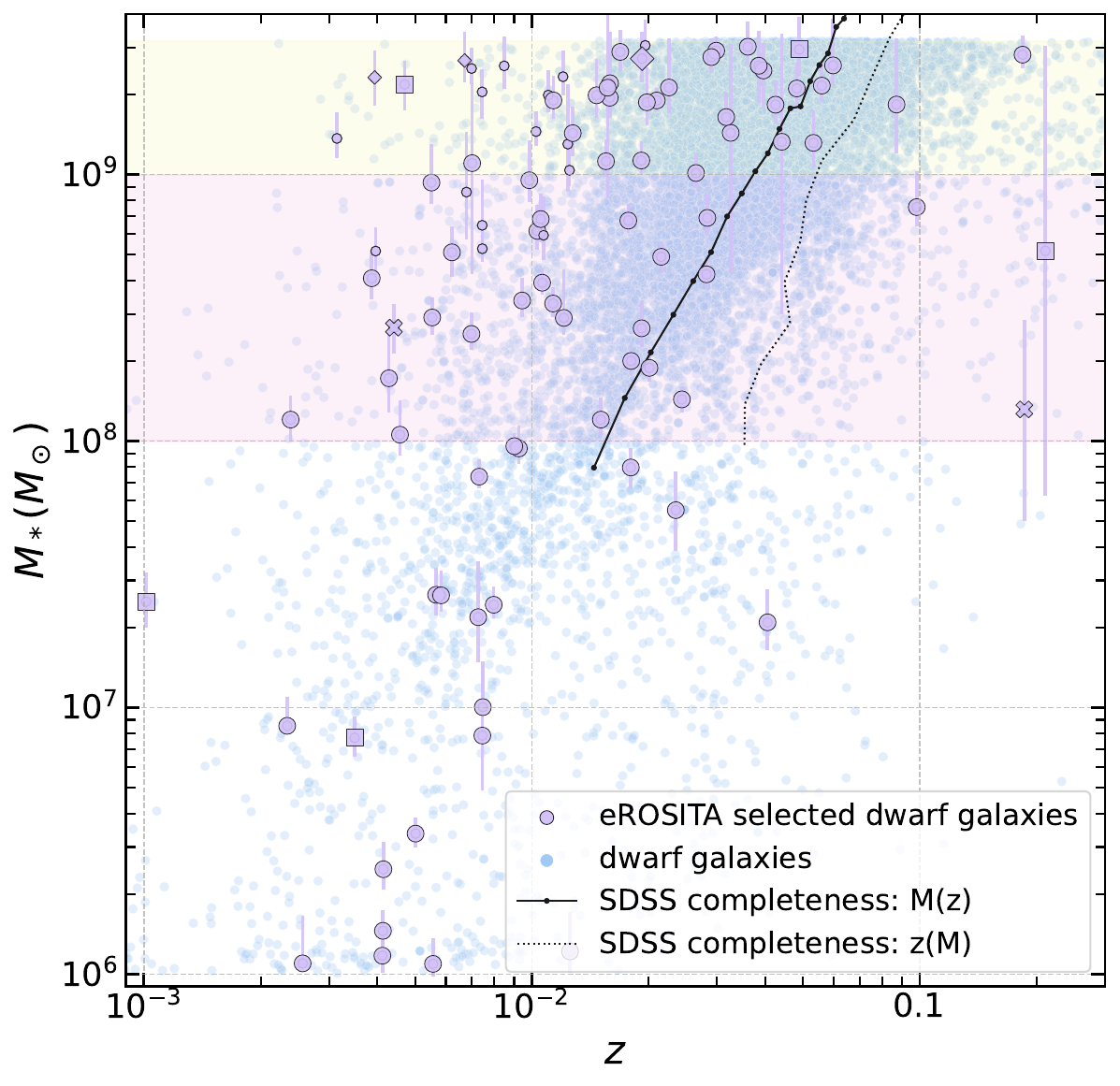}
    \caption[Dwarf galaxies stellar mass - redshift plane.]{Mass v. redshift plane. The approximate SDSS mass completeness limit in redshift bins is shown as a black curve, and SDSS redshift completeness in mass bins in the black dashed curve. We calculate the fraction of active dwarfs in stellar bins highlighted by horizontal stripes.}
    \label{fig:M_z}
\end{figure}

To control the possible completeness  of the dwarf galaxy content of  the MPA-JHU SDSS catalogue, we used three approaches. In the first approach, similar to \citealt{Birchall2022}, we define the redshift-dependent stellar-mass limit of SDSS $M(z)$ as a mass above which lies 90\% of galaxies in narrow redshift bins and use only galaxies with $M>M(z)$. In the second approach, we determine the mass-dependent redshift limit $z(M)$ below which lies 90\% of galaxies in a narrow stellar mass bin and use only galaxies with $z<z(M)$. In the third approach, we use all objects in a given mass bin. The  two selection curves are shown in Fig. \ref{fig:M_z} for $M_*>10^8 \msun$.

\begin{figure*}
    \includegraphics[width=\textwidth]{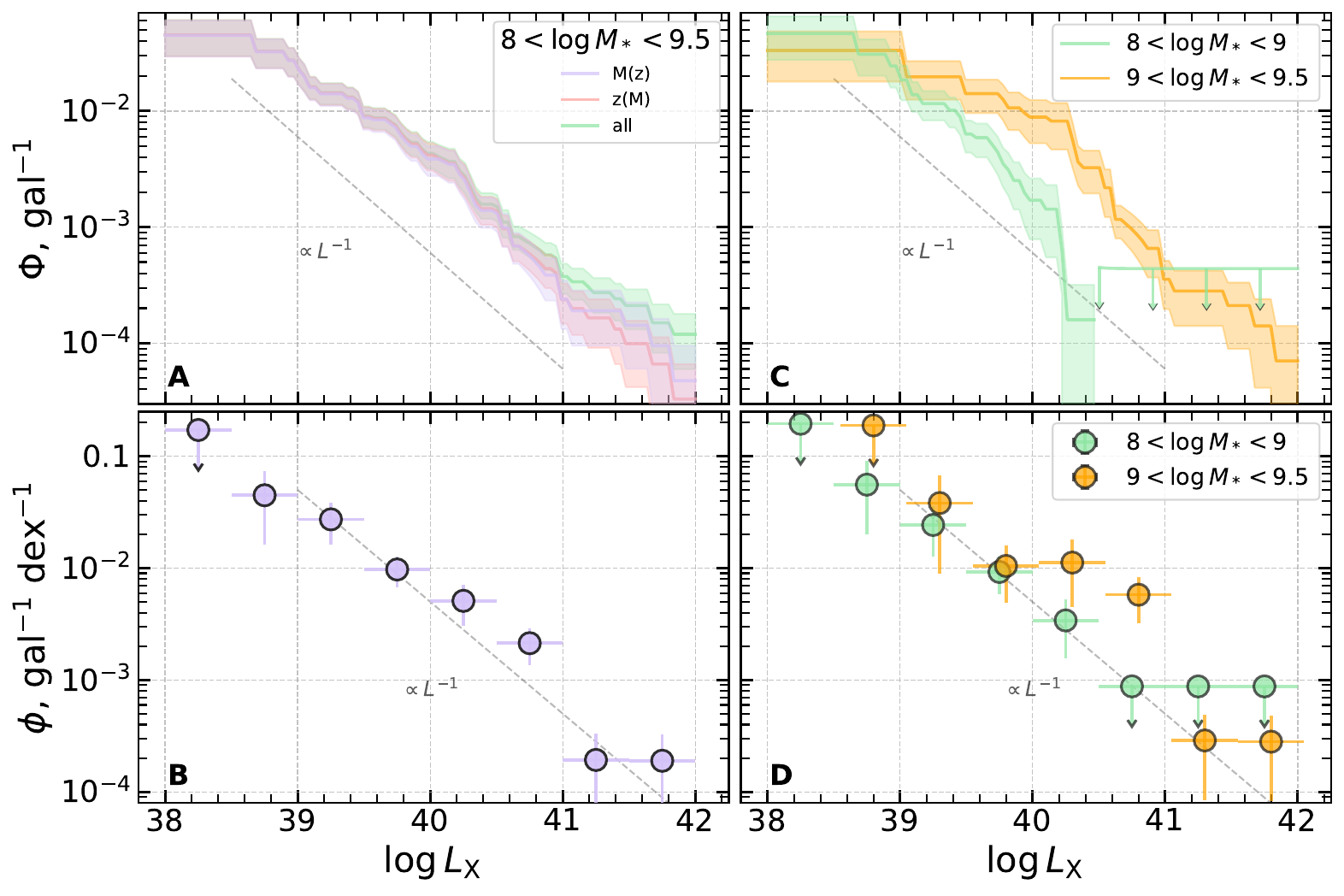}
    \caption[Dwarf galaxies differential and cumulative occupation fraction obtained with eROSITA.]{Results on the occupation fraction of AGN in dwarfs with $M_*>10^8 \msun$. Panel A shows the cumulative XLF for the three cases of calculated mass-completeness limits of SDSS. Panel B shows the differential XLF for one completeness case. Panels C and D show cumulative and differential XLF for two samples of dwarf galaxies: with masses below or above $10^9 \msun$. All panels show the power law with a slope of -1 to guide the eye. X-ray luminosities are in the 0.5-2 keV energy band.}
    \label{fig:agn_frac}
\end{figure*}

To compare these three approaches, we calculate the cumulative fraction of X-ray active  dwarf galaxies in a broad galaxy mass range $8<\log{M_*}<9.5$,  shown in Fig. \ref{fig:agn_frac},  panel A.  The three curves virtually coincide up to the luminosity of $\sim10^{40.5}$ and  diverge notably at $L_X\ge 10^{41.5}$, where Poissonian errors start to dominate. For further analysis, we chose to use the results of the first method of incompleteness correction, $M(z)$.

The differential XLF of dwarf galaxies is shown in panel B of Fig. \ref{fig:agn_frac}. Overall, the AGN fraction increases towards lower luminosities. The cumulative AGN fraction is $\sim2.24\pm0.59$ per cent at $L_{\rm X}>10^{39}$ and decreases to $\sim0.39\pm0.11$ per cent at $L_{\rm X}>10^{40}$.  There is evidence of flattening of the differential XLF  at the low luminosity end $L_{\rm X}<10^{38.5..39}$ (also seen in more narrow  mass bins, see below), but further data is needed to confirm  this finding. Some remaining systematic bias may be present at the low luminosity end of XLFs  because of the contamination by X-ray binaries.


To investigate the mass dependence of the   AGN fraction we separate the sample into two bins in stellar mass: $8<\log{M_*}<9$ and $9<\log{M_*}<9.5$. The results are shown in panels C and D of Fig. \ref{fig:agn_frac}. For luminosity $L_{\rm X}>10^{39}$ the cumulative AGN fraction for the low mass bin is $\sim1.85\pm0.61$ per cent, and $\sim3.32\pm1.53$ per cent for the high mass bin. For luminosity $L_{\rm X}>10^{40}$ the AGN fraction is $\sim0.17\pm0.09$ per cent (low mass bin) and $\sim0.89\pm0.36$ per cent (high mass bin). The difference in AGN fraction is statistically significant  for luminosity $L_{\rm X}>10^{40}$. This supports the picture of ubiquitous black hole occupation of massive galaxies which rises towards higher masses \citep{Aird2012,Aird2018}. The slope of XLF for the lower mass bin seems to be steeper than that of the higher mass.  We checked that making 'volume-limited' samples of dwarfs by restricting redshifts to be below the redshift of the lowest mass limit in a given mass bin does not change these results significantly. 

Differential XLF obtained in this paper is slightly steeper than that of \citealt{Birchall2020}\footnote{since they use band 8 of XMM for luminosities (0.2-12 keV), their luminosities are by $\sim0.5$ dex larger than ours} for the high mass bin, but overall it is consistent in magnitude. The cumulative AGN fraction for $L_{\rm X}>10^{40}$ for low and high mass bins from \citealt{Birchall2020} are roughly 0.5\% and 1\% respectively, slightly higher than the eROSITA results. The difference may arise from the selection criteria and our cleaning procedure, as discussed above. In this respect, we recall  that  \citeauthor{Birchall2020} excluded only 15 out of 101 initial matches. 

We did not find any statistically significant  redshift evolution of the AGN fraction.  


 MPA-JHU galaxies (and SDSS galaxies in general) are rather luminous and relatively close by, therefore the catalogue of dwarfs may not be representative of the entire dwarf galaxy populations in the local Universe, especially at higher redshifts.  We also found several obscured sources, against which eROSITA has strong bias due to the soft response. Obscured AGNs in dwarfs may or may not be a significant portion of the population, which we will likely miss. X-ray binaries may be problematic for a faint end of the luminosity function. One luminous X-ray binary is enough to produce $L_{\rm X}\sim10^{38..39}$ \ergps which we can deem as an AGN.

In the conclusion of this section, we mention an exciting  possibility to constrain the fraction of \textit{dormant} black holes in dwarf galaxies using the data of wide angle and long time span  surveys, like SRG/eROSITA all-sky survey. This possibility is given by the detection of TDEs in dwarf galaxies (for example discovered by eROSITA in ID 4296 in our sample, see discussion below). A TDE would be detected whether or not the host galaxy is classified as an AGN (in the case of ID 4296 it is a BPT-classified AGN). Detection of a TDE means that  a star was disrupted by a black hole, leaving no doubt that the centre of a galaxy is inhabited by a massive BH. ID 4296 is a dwarf with stellar mass $\sim3\times10^9 \msun$ and the corresponding BH mass of $\log M_{\rm BH}\sim5.8$, redshift is $z = 0.048$. Detection of just one TDE places a rather weak lower limit on the fraction of dormant black holes in dwarf galaxies, roughly one in $\sim40000$ galaxies. However, systematic wide-area sky surveys (also in optical, see \citealt{Yao2023}) will lead to an increase in the number of TDEs found in dwarf galaxies and will help to obtain more meaningful constraints.

\section{Individual objects}
\label{sect:indiv}

Based on the analysis of X-ray data, we selected  several sources with interesting properties worth some further discussion. Among them  a Tidal Disruption Event (TDE) candidate, a very soft and bright source but without significant variability, a luminous obscured AGN -- the most distant and luminous source in the sample and  an obscured and variable AGN. We also found several examples of X-ray active galaxy pairs and ultra-luminous X-ray source candidates.  The objects discussed in this section are listed in  Table \ref{tab:peculiar}.

\subsection{ID 4296 -- a possible Tidal Disruption Event in a dwarf galaxy}
\label{sect:indiv:tde}
Object 4296 (SRGe J023346.8-010129, SDSS J023346.93-010128.3) is peculiar. It is classified as an AGN (not LINER) in MPA-JHU catalogue (the top right source in Fig. \ref{fig:bpt}).  4296 has stellar mass $\sim3\times10^9 \msun$ and the corresponding BH mass of $\log M_{\rm BH}\sim5.8$ from the $M_*-M_{\rm BH}$ correlation. Redshift is $z = 0.048$ (distance of 203 Mpc). 

It has the most variable X-ray light curve (top left in Fig. \ref{fig:variability}) constantly rising through eRASS1..5, and the softest spectrum ($\Gamma=5.5\pm0.5$, Fig. \ref{fig:spectra}). The DESI LIS image is shown in Fig. \ref{fig:stamps} in the bottom right.  This object has a variable optical brightness and a transient event was discovered at that position in 2018 by Gaia Alerts Team (AT2018cqh\footnote{\url{https://www.wis-tns.org/object/2018cqh}}), but the event was not classified.

To track the evolution of the source's intensity we requested a Swift/XRT \citep{Burrows2005} Target of Opportunity observations. ToO request ID is 19205 with $4355$ s exposure in PC mode, performed on Aug 9, 2023. The data was processed using the standard Swift/XRT pipeline\footnote{\url{https://www.swift.ac.uk/analysis/xrt/}}. A source was detected in the given position with a significance over $3\sigma$, and its coordinates were pinpointed with \textsc{xrtcentroid} (position error $\sim5$"). The distance to the eROSITA position  is $1.2$", and to the SDSS dwarf is $2.5$". The source extraction region was centred on the source position and had a 20-pixel radius, whilst the background region is an annulus with radii of 25 and 60 pixels. Spectrum was grouped to 1 count per channel and fit with Cash statistic.

 Fit the co-added eROSITA data with the black body model gives the best-fitting temperature of $63\pm7  $ eV and the size of the emitting area of $R\sim1.9^{+0.8}_{-0.6}\times10^{11}$ cm for $203$ Mpc distance. No statistically significant temperature variations were detected between the individual sky surveys, although the statistical uncertainties in the first three surveys are fairly high. The Swift/XRT spectrum is similarly soft with the best-fitting temperature of $83\pm25$ eV, consistent with the  eROSITA value.  Both eROSITA and Swift/XRT data show evidence for the presence of a  non-thermal hard tail which we model as a power law with a fixed photon index $\Gamma=1.9$. For example, in eRASS 5 the C-statistic significantly improved from $31.67$ ($25$ dof) to $17.34$ ($24$ dof) with the added hard component, and it is responsible for $\sim10$ per cent of X-ray flux in the 0.3--2 keV energy band.  For estimating flux evolution we, however, use a simpler model without the hard component (\textsc{phabs*bbodyrad}). For each scan (and the XRT observation) we use individual best-fitting temperature and normalisation to estimate unabsorbed flux in the 0.3--2 keV energy range. 
We checked that the inclusion of the non-thermal component in the fit does not change the light curve discussed below significantly.

\begin{figure}
    \includegraphics[width=0.5\textwidth]{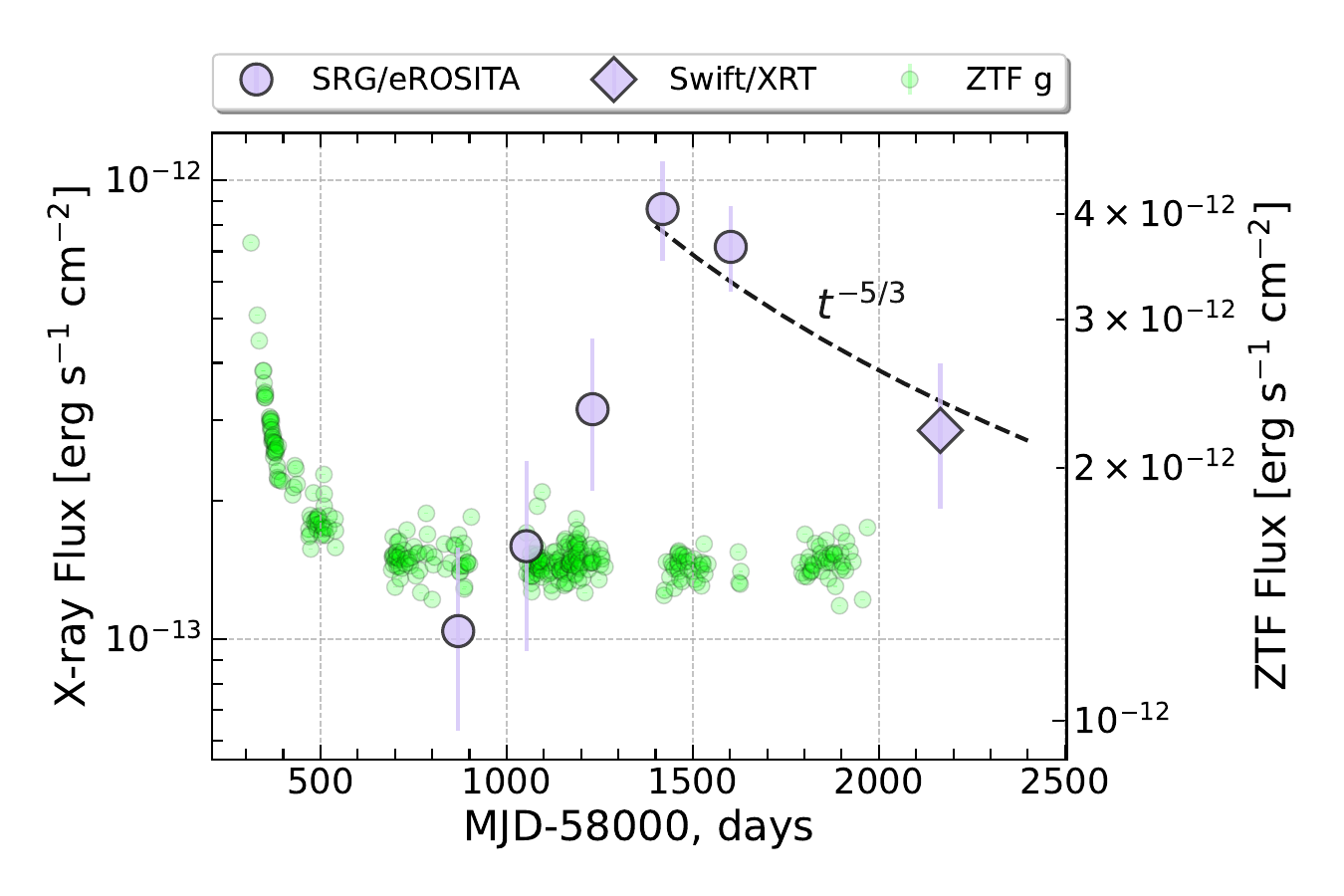}
    \caption[X-ray light curve of a TDE candidate in a dwarf galaxy.]{X-ray (0.3--2.0 keV) and ZTF light curve (g band) of ID 4296/AT2018cqh. The dashed line shows the expected $t^{-5/3}$ decline of X-ray luminosity expected from TDE.}
    \label{fig:variability-tde}
\end{figure}

Zwicky Transient Facility (ZTF, \citealt{Bellm2019}) light curve provides almost continuous data for this object from mid 2018\footnote{\url{https://ztf.snad.space/dr17/view/401310100001492},  \citep{Malanchev2023}; translated into flux using the reference ZTF g filter wavelength ($4783$ angstrom) and AB system zero point $4.76\times10^{-9}$ \ergpspcm ~ per angstrom, see \href{http://svo2.cab.inta-csic.es/theory/fps/index.php?id=Palomar/ZTF.g}{this url} } and is shown in Fig. \ref{fig:variability-tde} along with X-ray light curve. 

Optical data show a prominent flare which decayed to the quiescent state by the time of the first eROSITA observation. During this outburst, the galaxy became brighter by a least $1$ mag (i.e. by a factor of 2.5) for at least  200 days, meaning that at the peak of its light curve, the optical transient event was $\sim150$ per cent of the galaxy luminosity. X-ray light curve of eROSITA shows a gradual rise by a factor of 10 from eRASS 1 (MJD 58800) to eRASS 4 (MJD 59500) and hints at a decrease in brightness between eRASS 4 and 5. With a new data point from SWIFT, we find that in the course of the 1.5-year observation gap, the source brightness decreased by a factor of $\sim 2-3$.  The peak unabsorbed X-ray brightness corresponds to the $0.3-2$ keV luminosity of $\sim5.5\times10^{42}$ \ergps and is delayed with respect to optical emission by at least two years (assuming optical and X-ray events are related). The X-ray to optical ratio computed between peak values in each  band was $L_X/L_{\rm opt}\sim 0.3$.

Assuming the black hole mass quoted above ($\log M_{\rm BH}\sim5.8$) and ignoring bolometric correction, we estimate that the source was accreting on at least $\sim8$ per cent of Eddington level. A bolometric correction in the X-ray band using the best fit black body spectrum increases this number by a factor of $\sim 5$, to $\sim 40$ per cent.  The decay of the flux between eRASS4 and XRT observation roughly follows $\left(\frac{t-t_0}{\tau}\right)^{-5/3}$ trend expected for X-ray selected TDEs.  We also did not find the 'infrared echo' from the NeoWISE light curve\footnote{gathered with \href{https://github.com/HC-Hwang/wise_light_curves}{this python package}, see \citet{Hwang2020}} -- no increase in the W1 and W2 flux was detected.

The X-ray properties such as soft spectrum, large peak luminosity and flare-like light curve suggest that the source may be a Tidal Disruption event (TDE). 
The peak luminosity  is less than that of usual X-ray-selected TDEs ($L_{\rm X}\gtrsim10^{43}$ \ergps,  \citealt{Saxton2020, Sazonov2021}) but still noticeably larger than   other possibilities -- a nova, supernova, flaring star, X-ray binary (inc. ULX), see \citet{Zabludoff2021} for discussion of 'impostors'. The main alternative scenario is an AGN flare (see \citealt{Saxton2012, Auchettl2018, Neustadt2020, vanVelzen2021, vanVelzen2021b, Zabludoff2021}), especially given that the galaxy has an active nucleus according to SDSS spectroscopy from 2001. However, a typical AGN flare observed by eROSITA has an AGN-like hard X-ray spectrum \citep{Medvedev2022}. This is in line with the conclusion of  \citealt{Auchettl2018}, that X-ray selected TDEs have a drastically softer spectrum than  AGN at $z<2$, and moreover, the spectrum is usually much less absorbed. 
We conclude that  all the so far collected evidence strongly suggests that we caught a TDE.

Overall, several dozen TDEs active in X-rays have been identified so far, with only a handful of sources hosted by dwarf galaxies \citep{Maksym2013, Donato2014, Maksym2014,  Lin2017, He2021}. Optical TDEs are also found in dwarf galaxies, the first candidate was discovered recently in \citealt{Angus2022}.

There appears to be a dichotomy between X-ray bright and optically bright TDEs (Gilfanov et al., in preparation) with only some fraction, of the order of $\sim 20$ per cent, of X-ray bright TDEs also showing significant optical activity. Many of those active in both bands typically show a time delay between the peaks of the optical and X-ray emission. Although the full multi-wavelength picture is still emerging, the delay in ID 4296 seems to be much longer than observed in other events. 
It remains to be seen whether  the two-year-long delay is related to the properties of the host galaxy (dwarf with a low-mass black hole) or the parameters of the TDE itself.



The hard X-ray component may come either from the processes within the disrupted star remains, or be associated with the AGN in this dwarf (as per the SDSS BPT diagram, the galaxy was active long before eROSITA observations). eROSITA and XRT data are not sufficient  to reliably establish the variability of the hard component.


\subsection{ID 10487 -  AGN with a thermal spectrum in a dwarf galaxy }
\label{sect:indiv:soft_agn}
eROSITA ID 10487 is also one of the  three soft sources. It is located in a Seyfert galaxy RGG 123 at $z=0.0395$ (168 Mpc), with mass $\sim3\times10^9 \msun$.  \citealt{Chilingarian2018} has the black hole mass estimation of $M_{\rm BH}=1.11\pm0.07 \times 10^5 \msun$. There is a Chandra detection at the position of the galaxy with $L_{\rm X}=0.85\times10^{42}$ \ergps \citep{Greene2007, Dong2012, Baldassare2017} but without sufficient number of photons to constrain the spectrum -- only the photon index was estimated to be around 3 from the HR ratio \citep{Baldassare2017}. The Chandra data was taken in 2009-2010.

During eROSITA observations the source was quite bright (eROSITA detected 230 photons) and soft ($\Gamma=3\pm0.5$). The spectrum is well-fit with a \textsc{bbodyrad} model with interstellar absorption, eROSITA spectrum is shown in Fig. \ref{fig:indiv-spe-lc}. The temperature is $\sim160\pm15$ eV, the emission radius is $1\pm0.3\times10^{10}$ cm, the luminosity (0.3--8 keV) is $0.9\pm0.1\times10^{42}$ \ergps, consistent with the cited works (bolometric correction is small). With the  black hole mass quoted above, we estimate that the source is accreting at $\sim7$ per cent of Eddington luminosity. The multi-colour black body disc model \textsc{diskbb} also fits data well, with a temperature of $\sim240\pm30$ eV and a smaller radius (by a factor of $\sim2$, assuming inclination of $60\degr$). The source has a marginally significant ($\sim2.4\sigma$) hard power-law tail  -- spectral fit with only \textsc{diskbb} component gives C-stat 55.38 for 60 dof; adding a power law with fixed slope $\Gamma=1.9$ reduces  C-stat to 49.86 for 59 dof. In further discussion we fit the data without a hard component,  doing otherwise does not change the results within errors.  Neither eROSITA nor ZTF\footnote{\url{https://ztf.snad.space/dr17/view/480206200003099}} show evidence of brightness variability or flares.  The apparent stability of luminosity between Chandra and eROSITA observations, and the lack of faster optical and X-ray variability suggest that this source is not a TDE.  



The maximum temperature of the optically thick geometrically thin  accretion disc  is given by  \citep{Shakura1973} 
$$kT_{\rm max} = 11.5 \left(\frac{M_{\rm BH}}{10^8~M_\odot} \right)^{-1/4} {\dot{m}}^{1/4} \text{\rm (eV)}$$
where $\dot{m}$ is the accretion rate in terms of critical (Eddington)  accretion rate.  For the parameters of ID 10487 ($M_{\rm BH}=10^5$, $\dot{m}=0.07$), this formula gives the temperature of $kT=30$ eV, which is several times smaller than the values quoted above. The decrease in BH mass by about an order of magnitude and account for spectral hardening (due to the Compton scattering)  helps to explain the observed temperature.

To put this on a more quantitative footing we fit the data with the \textsc{grad} (General Relativistic Accretion Disk model around a Schwarzschild black hole, \citealt{Ebisawa1991}) model. Assuming the spectral hardening factor of $T_{\rm col}/T_{\rm eff}=2$ and inclination angle of $60\degr$ we obtain the best-fitting value of the black hole mass $M_{\rm BH}=1.0\pm0.3\times10^4$ $M_\odot$ and the mass accretion rate of $2.9\pm0.5\times10^{22}$ g s$^{-1}$,  giving the Eddington ratio of $1.5\pm0.5$ (accretion luminosity $L=2\pm0.3\times10^{42}$ \ergps). The fit provided an adequate description of the data with the C-stat 54.11 for 60 dof. 


To summarise, the soft thermal black body-like X-ray spectrum of this source suggests that we are observing  emission from the Shakura-Sunyaev accretion disk, which, thanks to the small mass of the central black hole emerges in the soft X-ray band. However,  the observed temperature appears to be too high and requires about $\sim 10$ times less massive black hole, than 
$\sim1\times10^5 \msun$  derived by \citet{Chilingarian2018} from optical spectroscopy. 

The observed soft spectrum of this source may resemble the soft excess  observed in AGN \citep{Gierlinski2004, Done2010, Done2012}. However, the soft excess  typically appears on top of the  hard power-law emission usually observed in AGN. For ID 10487, the hard power-law component seems to be subdominant.
All in all,  this scenario seems to be less likely and further investigation of this source in X-ray and optical bands is required to solve the puzzle of its soft X-ray spectrum.

The third soft source (ID 33862) has only 24 spectral counts, and precise spectral analysis is hampered. It is located in a radio galaxy FIRST J103837.1+443123 at $z=0.0124$ and has X-ray luminosity $\sim2\times10^{40}$ \ergps. The photon index is $\Gamma=4.5\pm1.5$ and black body temperature is $kT=110\pm10$ eV. The $\lxlxrbrat$ is only 1.2, but the soft X-ray spectrum cannot be explained by X-ray binaries. The contribution of the hot gas of ISM is not sufficient to explain the observed luminosity in the 0.5-2 keV energy range in all three soft sources.

\subsection{ID 45689,  11124 and 49241 -- peculiar absorbed sources in dwarf galaxies}
\label{sect:indiv:obscured_dwarf}

ID 45689 is a very luminous source. If the association between the SDSS dwarf  (z=0.21) and the eROSITA source is not a chance coincidence,  the dwarf has an obscured nucleus. eROSITA spectrum is fit with absorbed power law in the form of \textsc{phabs*zphabs*zpo}, yielding the intrinsic absorption (\textsc{zphabs}) of $N_{\rm H}=1.4^{+0.8}_{-0.6}\times10^{22}$ cm$^{-2}$. The photon index is fixed at 1.9. The unabsorbed luminosity is $L_{\rm X}=(1.4\pm0.4)\times 10^{44}$ \ergps in the 0.3--8 keV band.  

\begin{figure}
    \includegraphics[width=0.5\textwidth]{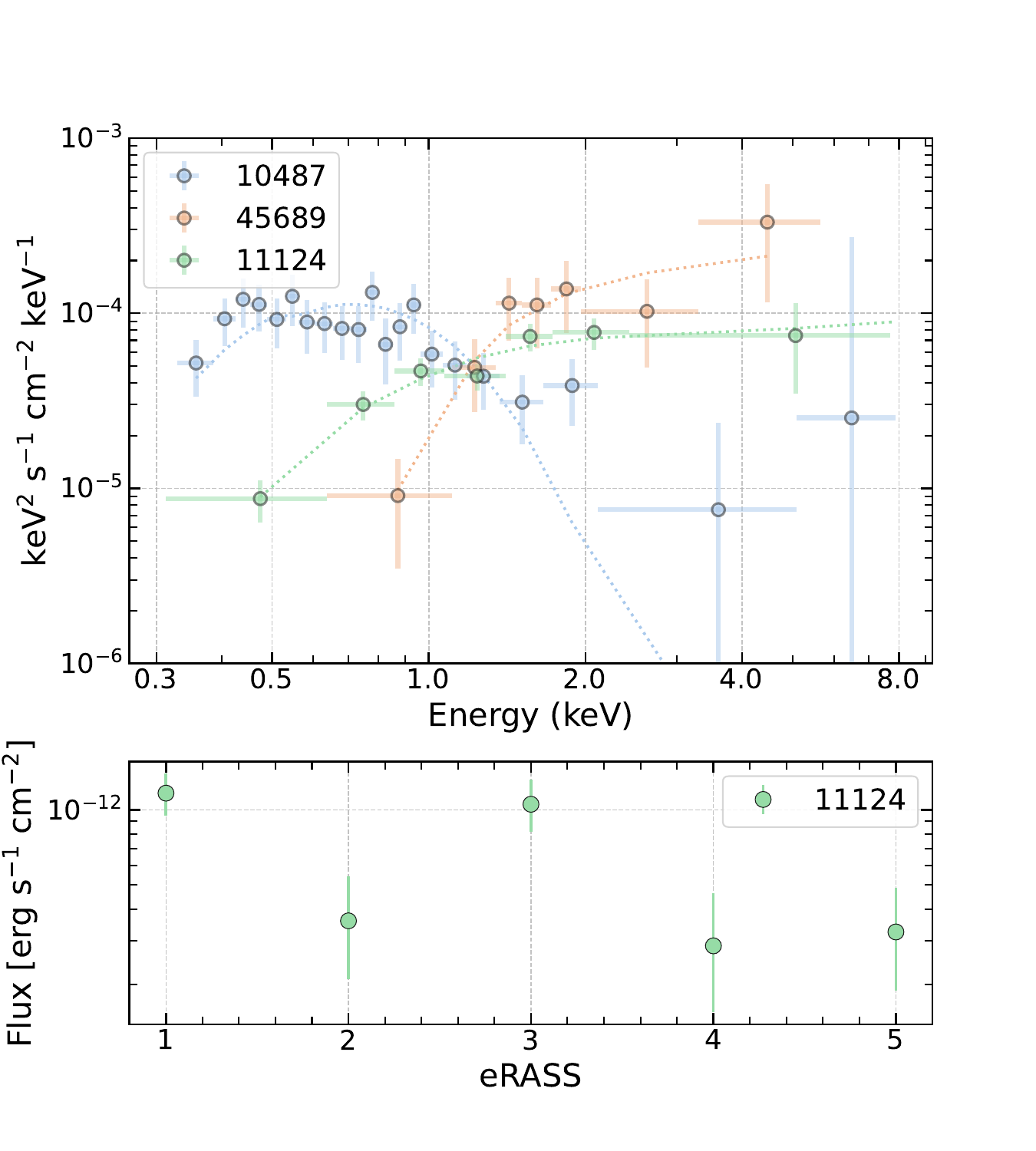}
    \caption[X-ray spectra and light curve of a few selected active dwarf galaxies]{X-ray spectra and light curve of a few selected sources discussed individually (ID 10487, 45689, 11124). In the top panel, unfolded eROSITA spectra are shown (rebinned for clarity). A power-law shape with $\Gamma=2.0$ in this plot would be a horizontal line. In the bottom panel, an X-ray light curve of ID 11124 is shown, as in Fig. \ref{fig:variability}.}
    \label{fig:indiv-spe-lc}
\end{figure}

To our knowledge, the largest reported luminosity of an AGN in a dwarf galaxy is  $10^{44}$ \ergps at $z=2.39$ found by \citealt{Mezcua2018} (their cid\_1192). Clearly, eROSITA found a quite rare object -- in \citealt{Mezcua2018, Zou2023}, none of the sources has luminosity above $10^{43}$ \ergps at $z<0.3$. The constraints on the bright end of the luminosity functions of active dwarfs may be important to assess the cosmic evolution of this population to constrain the scenarios of the AGN triggering and lifetime in low-mass black hole mass regime.  eROSITA spectrum for this source is shown in Fig. \ref{fig:indiv-spe-lc}.

ID 11124 is an obscured source (z=0.003493). It is fit in a similar fashion as ID 45689 above yielding intrinsic absorption of  $N_{\rm H}=0.18^{-0.05}_{+0.06}\times10^{22}$ cm$^{-2}$ and unabsorbed luminosity of $L_{\rm X}=(1.1\pm0.1)\times 10^{40}$ \ergps. The source seems to be variable:  in eRASS 1,3 it has the X-ray flux of $\sim 1.0\pm0.2\times10^{-12}$ \ergpspcm, and in eRASS 2,4,5 the flux is $0.3\pm0.15\times10^{-12}$ ($R=4.1$). The spectrum and light curve for this source are shown in Fig. \ref{fig:indiv-spe-lc}.

ID 49241 is located in a nearby dwarf galaxy NGC 4395. According to eROSITA data, ID 49241 shows a hard spectrum with a photon index $\Gamma=0.2\pm0.3$. The galaxy is known to have an X-ray source in a nucleus and is active \citep{Moran2005, Nardini2011}. X-ray analysis of archival data suggests the presence of variable absorption and changing spectral shape \citep{Kammoun2019}. 

\subsection{Galaxy pairs}
\label{sect:indiv:pairs}
There are several dwarf galaxies and/or eROSITA-SDSS matches excluded from the final list of X-ray active dwarfs which are located  close to another galaxy/galaxies.  It is important to note that we did not perform a \textit{systematic search} for galaxy pairs, but rather found several examples via visual inspection of DESI LIS images of each galaxy's surrounding. Those which we found will usually have Simbad type of 'Galaxy in a Pair of Galaxies', 'Galaxy in a Galaxy cluster' or 'Galaxy in a group of Galaxies'.   

\begin{figure*}
    \centering
    \includegraphics[width=0.7\textwidth]{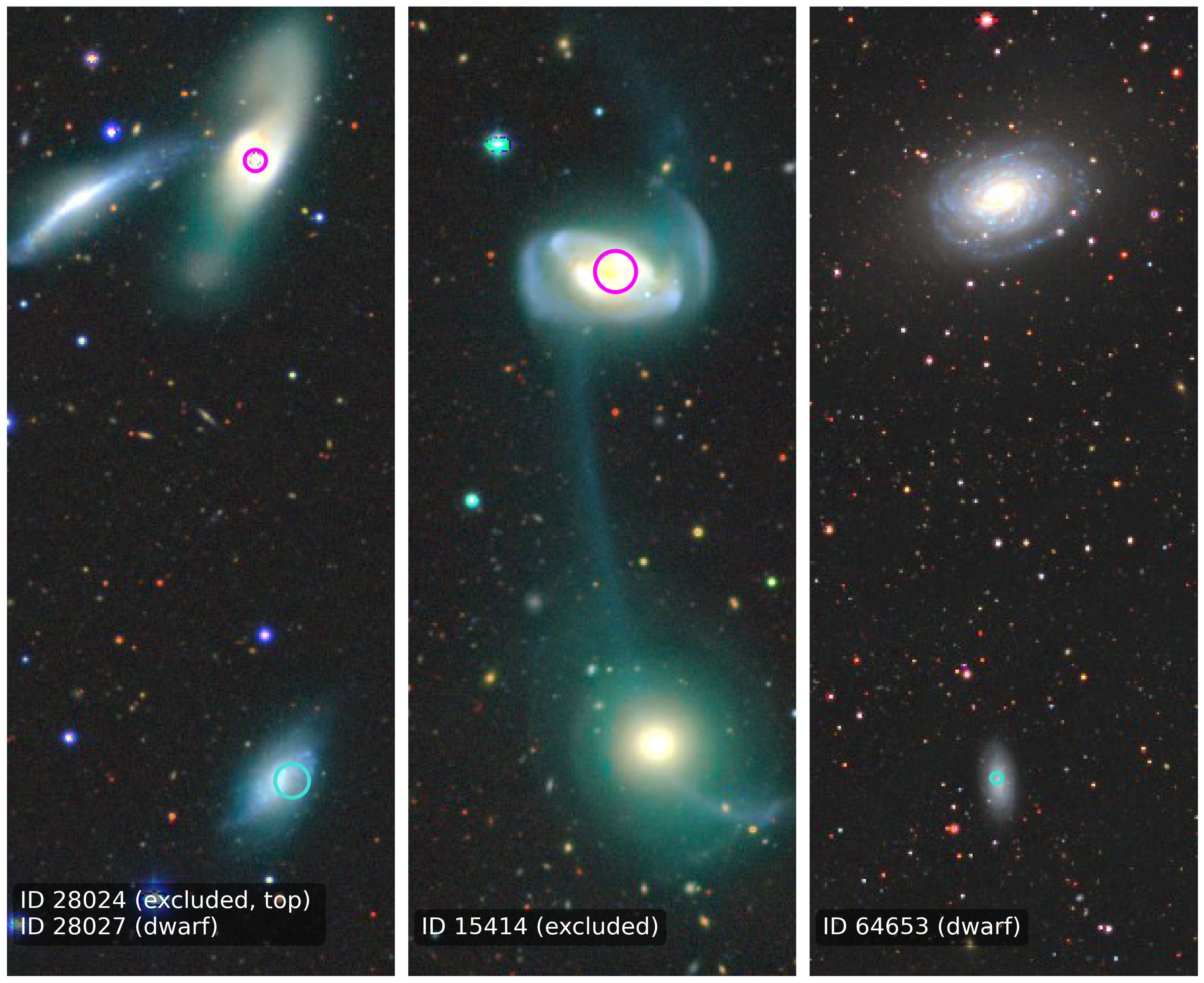}
    \caption[As Fig. \ref{fig:stamps}, but for galaxy pairs.]{As Fig. \ref{fig:stamps}, but for three examples of identified X-ray sources in  galaxy pairs, see sect. \ref{sect:indiv:pairs}. }
    \label{fig:stamps-pairs}
\end{figure*}

We show three examples in Fig. \ref{fig:stamps-pairs}. The galaxy separations reported below are projected separations and were calculated using the redshift of the Simbad galaxy associated with the eROSITA source.

The first example is the dwarf ID 28027 and excluded match  ID 28024 (likely a massive galaxy) which are located in galaxies UGC 4904 and NGC 2798 respectively, with the projected separation of $\sim40$ kpc. NGC 2798, in turn, interacts with galaxy NGC 2799 nearby. It may be the case that this pair is a rare example of 'dual AGN'. 

The second instance is excluded source 15414 in  galaxy NGC 5218 separated by $\sim50$ kpc from NGC 5216. The latter coincides with an eROSITA source not reported in this work. The third example, dwarf 64653, is identified as galaxy UGC 9925, a satellite of NGC 5962 \citep{Mao2021} some $\sim80$ kpc away. eROSITA sees a few sources near the core of NGC 5962, but they are not reported in this work.  

Other examples of pairs are ID 11387 (30 kpc projected separation, both galaxies active in X-rays), ID 64321 (6 kpc separation and both  active in X-rays), ID 34500 (6 kpc separation),  ID 25180 (150 kpc separation), ID 3474 (20 kpc separation). Excluded source ID 29963 is located in one of the interacting galaxy pair MCG+08-26-012 and 2MASX J14122652+4541254 (16 kpc separation). 

Some objects are located in a galaxy cluster or group, e.g. for dwarf galaxies those are ID 53431 (an X-ray source known before eROSITA), 16640, 2364, 53431 (a known X-ray source) and 32206; and for rejected matches ID 2356, 17032 and 23630. The latter is close to a gravitational lens candidate, but a follow-up is needed to confirm or exclude this association. 

Galaxy mergers are expected to play a significant role in AGN triggering and fuelling mechanisms and the corresponding growth of a massive black hole \citep{Mayer2010, Allevato2016}. Observational results imply that the AGN activity depends on the separation between pairs  \citep{Satyapal2014, Fu2018, Dougherty2023}, including the observation of rare dual-AGN \citep{McGurk2015}. There is evidence for the different behaviour observed in absorbed and unabsorbed X-ray AGNs \citep{Guainazzi2021} with decreasing separations. The  large-scale environment of AGN may play a role in determining the AGN triggering \citep{Hopkins2014, Allevato2016}, and  sources located in galaxy groups and clusters may aid further research in this direction if confirmed as AGN. The systematic search of AGN in interacting galaxies is within the scope of future research with eROSITA data.

\subsection{Ultra-luminous X-ray sources candidates}
\label{sect:indiv:ulx}

\begin{figure*}
    \centering
    \includegraphics[width=\textwidth]{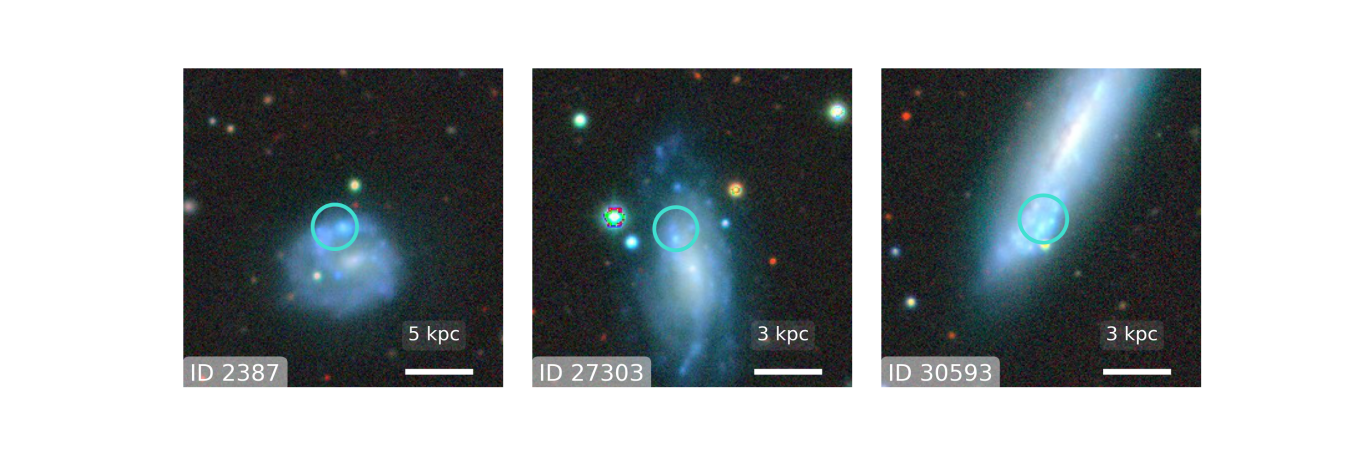}
    \caption[As Fig. \ref{fig:stamps}, but for ULX candidates]{As Fig. \ref{fig:stamps}, but for ULX candidates in the outskirts of dwarf galaxies, see sect. \ref{sect:indiv:ulx}.}
    \label{fig:stamps-ulx}
\end{figure*}

During the visual inspection of sources, we found three cases when eROSITA clearly detects off-centre X-ray emission in a dwarf galaxy.  This happens because the MPA-JHU catalogue sometimes includes  several fibre positions for one galaxy, and the eROSITA source may be matched with one of them.  The sources in question are ID 2387, 27303 and 30593, their optical images are shown in Fig. \ref{fig:stamps-ulx}. The sources are removed from the active dwarf catalogue because they are off-centred (sect. \ref{sect:catalog:cross-match}).  Note that ULXs may be also responsible for \textit{nuclear} X-ray emission of some dwarf galaxies, but these cases are quite difficult to discern from a genuine AGN based on eROSITA data only. None the less, based on the expected frequency of ULXs, we do not anticipate that they dominate our sample of 99 active dwarfs.

X-ray spectra of these sources are consistent with the canonical AGN spectrum (see sect. \ref{sect:catalog:xray}). ID 2387 and 27303 have luminosities of $\sim5\times 10^{39}$ \ergps, and 30593 has a luminosity of $\sim10^{40}$ \ergps. 

Ultra-luminous X-ray sources, ULX,   \citep{Fabbiano1989, Swartz2004, Kaaret2017, Fabrika2021, King2023} is a kind of an X-ray object which has luminosity over $10^{39}$ \ergps (an Eddington luminosity of $10\msun$ object) and is not (usually) located at the centre of its host galaxy. About 1800 ULX candidates are known so far \citep{Walton2022}. It is suggested that intermediate-mass black holes (IMBH) or super-critically accreting pulsars may be responsible for the phenomena.  Even though our search is not  a substitute for a systematic  ULX survey, the three ULXs found may present a useful sample for the population studies of this kind of object. It is observed that the ULX populations depend on the galaxy morphological type, metallicity, mass and star formation rate \citep{Kovlakas2020}, with dwarfs hosting more ULXs than expected from their SFR \citep{Walton2011, Plotkin2014, Kovlakas2020}. Further discussion is much beyond the scope of this paper. None of our three ULX candidates are found in the catalogue of 1800 ULX candidates of  \citeauthor{Walton2022}.

\begin{table}
\begin{tabular}{ll}
\multicolumn{1}{c|}{ID}    & Peculiarity     \\ \hline
\multicolumn{2}{c}{Dwarf galaxies}          \\
\multicolumn{1}{l|}{4296}  & TDE cand. (sect. \ref{sect:indiv:tde})      \\
\multicolumn{1}{l|}{10487} & Soft state AGN (sect. \ref{sect:indiv:soft_agn}) \\
\multicolumn{1}{l|}{45689} & Obscured AGN (sect. \ref{sect:indiv:obscured_dwarf})  \\
\multicolumn{1}{l|}{28027} & Gal. pair (with 28024)      \\ 
\multicolumn{1}{l|}{11387} & Gal. pair     \\ 
\multicolumn{1}{l|}{64321} & Gal. pair     \\ 
\multicolumn{1}{l|}{34500} & Gal. pair     \\ 
\multicolumn{1}{l|}{64653} & Gal. pair     \\ 
\multicolumn{1}{l|}{25180} & Gal. pair     \\ 
\multicolumn{1}{l|}{3474} & Gal. pair     \\ 
\multicolumn{1}{l|}{53431} & In group     \\ 
\multicolumn{1}{l|}{32206} & In group     \\ 
\multicolumn{1}{l|}{2364} & In group     \\ 
\multicolumn{1}{l|}{16640} & In cluster     \\ 
\multicolumn{2}{c}{Excluded sources}        \\ \hline
\multicolumn{1}{l|}{2387}  & ULX cand. (sect. \ref{sect:indiv:ulx})     \\
\multicolumn{1}{l|}{27303} & ULX cand.  (sect. \ref{sect:indiv:ulx})    \\
\multicolumn{1}{l|}{30539} & ULX cand.  (sect. \ref{sect:indiv:ulx})    \\ 
\multicolumn{1}{l|}{28024} & Gal. pair (with 28027)      \\ 
\multicolumn{1}{l|}{15414} & Gal. pair     \\ 
\multicolumn{1}{l|}{29963} & Gal. pair     \\             
\multicolumn{1}{l|}{23630} & Grav lens.?     \\ 
\multicolumn{1}{l|}{2356} & In group     \\ 
\multicolumn{1}{l|}{17032} & In group     \\ 
\end{tabular}
\caption[List of individual sources discussed in detail.]{List of individually discussed/mentioned sources (sect. \ref{sect:indiv})}
\label{tab:peculiar}
\end{table}

\section{Conclusion}
\label{sect:conclusion}

We performed a systematic and rigorous search for X-ray signatures of accreting massive black holes in dwarf galaxies from the MPA-JHU catalogue (based on SDSS data). We used the data from the SRG/eROSITA all-sky survey in the Eastern galactic hemisphere.  In total, we found 178 matches between the eROSITA source catalogue and low mass galaxies ($M_*<10^{9.5}\msun$) in the MPA-JHU  catalogue. We estimate the number of spurious matches as $\sim17$ per cent. The initial matches between X-ray sources and dwarf galaxies were conservatively cleaned to achieve high purity of the final catalogue of X-ray active dwarf galaxies, including the removal of quasars and massive galaxies (sect. \ref{sect:catalog:cross-match}).  As a result, we presented a catalogue of 99 dwarf galaxies with nuclear X-ray activity. Only 14 sources were reported as possible active dwarf candidates in previous studies (sect. \ref{sect:results:known_dwarfs}). 

We performed a detailed analysis of  the X-ray properties of objects, including spectral and variability studies (sect. \ref{sect:catalog:xray}). We estimate the contribution from X-ray binaries and the hot ISM gas to the observed X-ray emission to find that the majority of sources (82/99)  are in fact strong candidates for being an AGN in the dwarf galaxy (sect. \ref{sect:catalog:xrb}).

We discussed the host galaxy properties of selected sources (sect. \ref{sect:results}). Emission line diagnostics proposes that almost all sources are star-forming galaxies, emphasising the importance of X-ray surveys for discovering low-luminosity AGN in low-mass galaxies (sect. \ref{sect:results:optical:bpt}). Assuming a relationship between galaxy stellar mass and the mass of the central black hole we estimate that the bulk of sources are accreting at around 1 per cent of the critical accretion rate (sect. \ref{sect:results:optical:eddington}).

We estimate the fraction of dwarfs with active nuclei (sect. \ref{sect:results:optical:fraction}).  We find that the occupation fraction of dwarfs with AGN falls with increasing X-ray luminosity and spans from $\sim0.01-2$ per cent. The AGN fraction increases with host mass, at least for luminosities above $10^{40}$ \ergps, from $\sim0.2$ per cent to $\sim1$ per cent between $8<\log{M_*}<9$ and $9<\log{M_*}<9.5$. We discuss possible selection effects affecting the measured occupation fraction. The measurements of AGN fraction impose a lower limit on the occupation fraction of dwarf galaxies with central black holes. 

We present the catalogues with optical and X-ray properties of all 99 X-ray active dwarf galaxies (Appendix  \ref{appendix:catalog}, Table \ref{tab:dwarfs}). We also present the list of matches which were rejected in the course of the cleaning procedure (Appendix \ref{appendix:excluded}).

We serendipitously discovered several interesting sources. A prominent object is a transient eROSITA source coinciding with a known optical transient. The X-ray spectral shape and variability suggest that this source may be a tidal disruption event hosted by an active dwarf galaxy at $z=0.048$. This is a novel addition to a few known TDEs in dwarf galaxies selected in X-rays and optical wavelengths. 

Another source with a soft X-ray spectrum is fit with thermal emission with a black body temperature of 160 eV. The source spectrum suggests a disc-dominated state of an accreting massive black hole, however, the X-ray spectral fit requires almost an order of magnitude less massive black hole, than inferred from optical spectroscopy.

A few sources with hard spectra are found to be of an obscured nature. One of those is a very luminous and relatively close AGN rivalling the luminosity record holder in dwarf galaxies. Another obscured source is variable in X-rays on the time scale of 6 months.

We found an array of dwarf galaxies (as well as eROSITA sources from the excluded list of galaxies) which are located in galaxy pairs, groups and clusters.  The found projected separations span from $\sim5$ to $\sim150$ kpc and contain four examples when the X-ray source is visible in both galaxies in a pair. 6 sources are located in compact galaxy groups or clusters of galaxies. In future, a complete eROSITA sample of this kind of object may give important information towards the AGN triggering mechanisms. 

We report three ULX (Ultra-luminous X-ray sources) candidates found in three dwarf galaxies. These sources will allow us to better populate  the low-mass regime of the vast ULX population and  better study the environmental dependencies of the occurrence of this type of object.


\section*{Acknowledgements}
We thank the anonymous referee for useful and constructive comments and suggestions which helped to improve the presentation of our results.
Sergei Bykov acknowledges support from and participation in the International Max-Planck Research School (IMPRS) on Astrophysics at the Ludwig-Maximilians University of Munich (LMU).

This work is based on observations with the eROSITA telescope onboard the SRG observatory. The SRG observatory was built by Roskosmos in the interests of the Russian Academy of Sciences represented by its Space Research Institute (IKI) in the framework of the Russian Federal Space Program, with the participation of the Deutsches Zentrum für Luft-und Raumfahrt (DLR). The SRG/eROSITA X-ray telescope was built by a consortium of German Institutes led by MPE, and supported by DLR.  The SRG spacecraft was designed, built, launched and is operated by the Lavochkin Association and its subcontractors. The science data are downlinked via the Deep Space Network Antennae in Bear Lakes, Ussurijsk, and Baykonur, funded by Roskosmos. The eROSITA data used in this work were processed using the eSASS software system developed by the German eROSITA consortium and proprietary data reduction and analysis software developed by the Russian eROSITA Consortium. This work was partially supported by grant 21-12-00343 from the Russian Science Foundation.

This work used the data of the Sloan Digital Sky Survey. Funding for the Sloan Digital Sky Survey V has been provided by the Alfred P. Sloan Foundation, the Heising-Simons Foundation, the National Science Foundation, and the Participating Institutions. SDSS acknowledges support and resources from the Center for High-Performance Computing at the University of Utah. The SDSS web site is \url{www.sdss.org}. SDSS is managed by the Astrophysical Research Consortium for the Participating Institutions of the SDSS Collaboration, including the Carnegie Institution for Science, Chilean National Time Allocation Committee (CNTAC) ratified researchers, the Gotham Participation Group, Harvard University, Heidelberg University, The Johns Hopkins University, L’Ecole polytechnique federale de Lausanne (EPFL), Leibniz-Institut fuer Astrophysik Potsdam (AIP), Max-Planck-Institut fuer Astronomie (MPIA Heidelberg), Max-Planck-Institut fuer Extraterrestrische Physik (MPE), Nanjing University, National Astronomical Observatories of China (NAOC), New Mexico State University, The Ohio State University, Pennsylvania State University, Smithsonian Astrophysical Observatory, Space Telescope Science Institute (STScI), the Stellar Astrophysics Participation Group, Universidad Nacional Aut\'{o}noma de M\'{e}xico, University of Arizona, University of Colorado Boulder, University of Illinois at Urbana-Champaign, University of Toronto, University of Utah, University of Virginia, Yale University, and Yunnan University.

The authors are grateful to the Swift observatory team for approving a ToO observation 19205 of a TDE candidate. This work made use of data supplied by the UK Swift Science Data Centre at the University of Leicester

Software: AstroPy \citep{astropy:2018}, Pandas\citep{reback2020pandas},  NumPy \citep{Harris2020}, Matplotlib \citep{Hunter2007}, HEALPy \citep{Zonca2019}. 
This research has made use of the SIMBAD database,
operated at CDS, Strasbourg, France.

\section*{Data Availability Statement}
SDSS data is available through the links in sect. \ref{sect:data:galaxy}.
The catalogues presented in this article are available as online supplementary material.
Swift/XRT ToO data for ID 4296 is available as \href{https://swift.gsfc.nasa.gov/cgi-bin/sdc/list?seq=sw00016191001.017}{this url}.
SRG/eROSITA data on  the sources published in this paper can be made available upon a reasonable request.
The catalogue of active dwarf candidates presented here will be made publicly available via the VizieR\footnote{\url{https://vizier.cds.unistra.fr/viz-bin/VizieR}} system after  the publication of this work.



\bibliographystyle{mnras}
\bibliography{dwarfs} 



\appendix
\section{Catalogue description}
\subsection{Catalogue of dwarfs}
\label{appendix:catalog}

In the accompanying text file \textsc{dwarfs.csv} one finds the catalogue of 99 X-ray selected dwarfs. It has 99 rows and 67 columns. Columns include the relevant identification information; positional information (optical/X-ray coordinates, separations); host galaxy properties - redshift, mass, star formation rate, expected X-ray binaries population, emission line intensities relevant for the BPT diagram; object identification according to SIMBAD; X-ray information - detection likelihood, positional errors, source counts, X-ray luminosities in several bands, variability parameter, fiducial spectral model. When possible, errors are provided for either 68\% or 90\% intervals. The header of the file explains every column.

Table \ref{tab:dwarfs} presents a quick-look subset of columns for all 82 dwarfs with $\lxlxrbrat>3$. Namely, optical and X-ray coordinates, host properties (redshift, stellar mass, star formation rate), X-ray luminosity, and its ratio to the expected XRB contribution.

In the attached pdf file \textsc{dwarfs\_stamps.pdf} an image of every dwarf can be found (as in Fig. \ref{fig:stamps}).

\subsection{Catalogue of excluded sources}
\label{appendix:excluded}

In the accompanying text file \textsc{rejected.csv} one may find the catalogue of 79 eROSITA sources which were cleaned in sect. \ref{sect:catalog:cross-match} with the appropriate rejection reason. The structure of the catalogue is similar to that of 99 dwarfs but without XRB population, emission line intensities, X-ray luminosities (only fluxes) and variability parameter\footnote{X-ray spectral analysis was not performed for excluded sources}.

\begin{figure*}
    \centering
    \includegraphics[width=0.8\hsize]{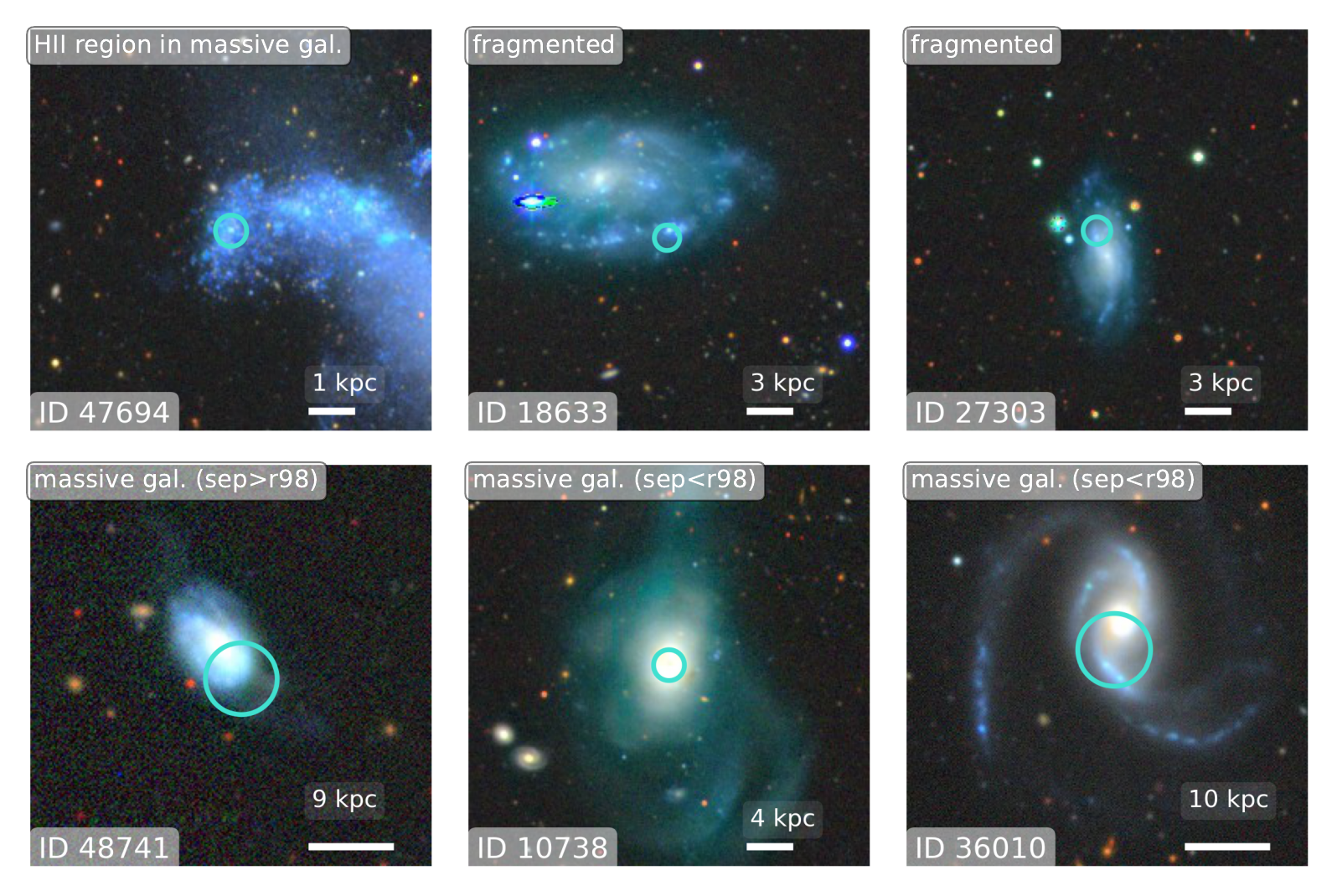}
    \caption[As Fig. \ref{fig:stamps}, but for excluded sources.]{As Fig. \ref{fig:stamps}, but for galaxies which were cleaned during the cross-match process. Each galaxy has an ID on the bottom left, and the reason for exclusion on the top left. All except 10738 and 36010 were removed after visual inspection, and 10738 and 36010 were removed automatically as per massive galaxy within $r_{98}$ criterion. 27303 is a notable example of a dwarf galaxy with an eROSITA source in the outskirts.}
    \label{fig:stamps-bad}
\end{figure*}

Some examples of excluded sources are shown in Fig. \ref{fig:stamps-bad} with the reason for exclusion marked. All images can be found in the attached pdf file \textsc{excluded\_stamps.pdf} (as in Fig. \ref{fig:stamps}).

\onecolumn
\begin{landscape}
{
\fontsize{7}{9}\selectfont
\begin{longtable}{llllllllllll}
\caption{SRG/eROSITA--SDSS catalogue of dwarf galaxies with nuclear X-ray activity.} \label{tab:dwarfs} \\
\toprule
ID$^{(1)}$ & RA$^{(2)}$ & DEC$^{(3)}$ & Simbad$^{(4)}$ & $z$$^{(5)}$ & ID$_{\rm X}$$^{(6)}$ & RA$_{\rm X}$$^{(7)}$ & DEC$_{\rm X}$$^{(8)}$ & $\sigma_{\rm X}$$^{(9)}$ & $M_*, M_{\odot}$$^{(10)}$ & $L_{\rm X, 0.3-8}$ erg s$^{-1}$$^{(11)}$ & $\frac{L_{\rm X, 0.5-8}}{L_{\rm X, XRB}}$$^{(12)}$ \\
\midrule
\endfirsthead
\caption[]{SRG/eROSITA--SDSS catalogue of dwarf galaxies with nuclear X-ray activity.} \\
\toprule
ID$^{(1)}$ & RA$^{(2)}$ & DEC$^{(3)}$ & Simbad$^{(4)}$ & $z$$^{(5)}$ & ID$_{\rm X}$$^{(6)}$ & RA$_{\rm X}$$^{(7)}$ & DEC$_{\rm X}$$^{(8)}$ & $\sigma_{\rm X}$$^{(9)}$ & $M_*, M_{\odot}$$^{(10)}$ & $L_{\rm X, 0.3-8}$ erg s$^{-1}$$^{(11)}$ & $\frac{L_{\rm X, 0.5-8}}{L_{\rm X, XRB}}$$^{(12)}$ \\
\midrule
\endhead
\midrule
\multicolumn{12}{r}{Continued on next page} \\
\midrule
\endfoot
\bottomrule
\endlastfoot
2349 & 253.88911 & 63.24213 & NGC  6275 & 0.023 & SRGe J165533.6+631432 & 253.88992 & 63.24216 & 5.8 & $2.1^{+1.1}_{-0.4}\times10^{9}$ & $3.4\pm0.7\times10^{40}$ & 3.1 \\
2364 & 253.90256 & 64.01973 & SDSSCGB 51726.2 & 0.023 & SRGe J165536.2+640107 & 253.90074 & 64.01858 & 5.6 & $5.5^{+2.2}_{-1.6}\times10^{7}$ & $1.7\pm0.7\times10^{40}$ & 302.4 \\
2506 & 262.31793 & 59.56275 & SDSS J172916.31+593346.0 & 0.018 & SRGe J172915.7+593345 & 262.31561 & 59.5625 & 6.4 & $2.0^{+0.34}_{-0.24}\times10^{8}$ & $1.3\pm0.4\times10^{40}$ & 19.9 \\
2556 & 262.07526 & 60.94234 & UGC 10880 & 0.013 & SRGe J172817.5+605630 & 262.07297 & 60.9416 & 8.5 & $1.22^{+0.5}_{-0.21}\times10^{6}$ & $5.8\pm1.7\times10^{39}$ & $5.0\times10^{3}$ \\
2595 & 257.75494 & 56.90007 & SDSS J171101.19+565400.2 & 0.028 & SRGe J171101.4+565401 & 257.75573 & 56.90038 & 7.2 & $6.9^{+1.6}_{-0.9}\times10^{8}$ & $2.3\pm1.1\times10^{40}$ & 24.6 \\
3474 & 7.71736 & 0.52872 & LEDA    1881 & 0.019 & SRGe J003051.9+003149 & 7.71617 & 0.53035 & 10.8 & $1.13^{+0.21}_{-0.13}\times10^{9}$ & $2.2\pm1.5\times10^{40}$ & 7.9 \\
4296 & 38.44556 & -1.02454 & SDSS J023346.93-010128.3 & 0.049 & SRGe J023346.8-010129 & 38.44505 & -1.02478 & 5.0 & $2.9^{+1.0}_{-0.6}\times10^{9}$ & $2.01\pm0.28\times10^{42}$ & 385.6 \\
5545 & 29.86125 & 14.83424 & LEDA 1468320 & $7.0\times10^{-3}$ & SRGe J015926.8+145002 & 29.86185 & 14.83388 & 5.4 & $2.52^{+0.5}_{-0.33}\times10^{8}$ & $4.1\pm0.9\times10^{40}$ & 123.5 \\
7394 & 195.00429 & 66.87216 & - & 0.048 & SRGe J130000.5+665216 & 195.00222 & 66.871 & 5.4 & $2.1^{+0.5}_{-0.35}\times10^{9}$ & $1.14\pm0.17\times10^{42}$ & 543.2 \\
9225 & 143.50993 & 55.23978 & Mrk  116B & $2.6\times10^{-3}$ & SRGe J093402.0+551427 & 143.5085 & 55.24084 & 5.6 & $1.098^{+0.6}_{-0.03}\times10^{6}$ & $5.0\pm0.9\times10^{39}$ & 11.9 \\
10487 & 233.60663 & 4.1352 & [RGG2013] 123 & 0.04 & SRGe J153425.5+040806 & 233.60638 & 4.13505 & 5.0 & $2.5\pm0.7\times10^{9}$ & $9.3\pm1.2\times10^{41}$ & 222.0 \\
10619 & 173.33403 & 63.27942 & UGC  6534 & $4.1\times10^{-3}$ & SRGe J113320.8+631646 & 173.33667 & 63.2795 & 11.3 & $2.5^{+0.7}_{-0.4}\times10^{6}$ & $2.5\pm1.1\times10^{39}$ & $1.9\times10^{3}$ \\
11048 & 233.80957 & 57.51482 & 2MASX J15351422+5730529 & 0.01 & SRGe J153513.7+573053 & 233.80694 & 57.51461 & 6.1 & $6.1^{+1.6}_{-0.9}\times10^{8}$ & $7.6\pm2.3\times10^{39}$ & 11.8 \\
11092 & 233.36633 & 53.51831 & SDSS J153327.91+533105.9 & 0.087 & SRGe J153326.8+533105 & 233.36171 & 53.51814 & 10.1 & $1.8^{+1.4}_{-0.6}\times10^{9}$ & $4.2\pm1.8\times10^{41}$ & 21.1 \\
11124 & 234.2674 & 55.26406 & SDSS J153704.18+551550.5 & $3.5\times10^{-3}$ & SRGe J153704.5+551550 & 234.26857 & 55.26376 & 5.0 & $7.7^{+1.5}_{-1.2}\times10^{6}$ & $1.79\pm0.17\times10^{40}$ & 113.8 \\
11195 & 237.02917 & 55.719 & NVSS J154806+554305 & 0.04 & SRGe J154806.7+554307 & 237.02774 & 55.71859 & 10.2 & $2.1^{+0.7}_{-0.4}\times10^{7}$ & $9.0\pm3.5\times10^{40}$ & 255.1 \\
11289 & 242.79799 & 48.33444 & 2MASX J16111153+4820036 & $9.4\times10^{-3}$ & SRGe J161111.5+482005 & 242.79792 & 48.33472 & 6.7 & $3.4^{+0.7}_{-0.4}\times10^{8}$ & $6.9\pm1.9\times10^{39}$ & 5.0 \\
11291 & 243.02216 & 48.8183 & - & 0.02 & SRGe J161205.4+484905 & 243.02259 & 48.8181 & 7.0 & $1.88^{+0.31}_{-0.21}\times10^{8}$ & $2.7\pm0.9\times10^{40}$ & 44.3 \\
11530 & 249.91254 & 43.29555 & MCG+07-34-128 & 0.016 & SRGe J163938.9+431741 & 249.91211 & 43.29459 & 12.2 & $1.94^{+0.4}_{-0.25}\times10^{9}$ & $1.5\pm0.6\times10^{40}$ & 5.2 \\
11568 & 250.23807 & 44.94012 & [VV2003c] J164057.1+445624 & 0.018 & SRGe J164056.8+445620 & 250.23652 & 44.93901 & 8.7 & $8.0^{+1.5}_{-1.3}\times10^{7}$ & $1.2\pm0.6\times10^{40}$ & 29.9 \\
11569 & 250.68242 & 45.71156 & LEDA 2269311 & 0.015 & SRGe J164243.4+454244 & 250.68074 & 45.71209 & 10.1 & $1.2^{+0.26}_{-0.16}\times10^{8}$ & $2.1\pm0.6\times10^{40}$ & 39.5 \\
11622 & 254.62193 & 40.95593 & SDSS J165829.26+405721.5 & 0.028 & SRGe J165829.7+405721 & 254.62361 & 40.95597 & 12.5 & $4.2^{+0.6}_{-0.5}\times10^{8}$ & $8.6\pm2.2\times10^{40}$ & 69.2 \\
11958 & 356.44275 & -8.78514 & LEDA  999228 & 0.056 & SRGe J234545.4-084707 & 356.43932 & -8.7854 & 12.6 & $2.16^{+0.5}_{-0.28}\times10^{9}$ & $3.6\pm2.0\times10^{41}$ & 69.1 \\
12509 & 26.14118 & -8.7731 & MCG-02-05-039 & 0.03 & SRGe J014433.6-084630 & 26.14016 & -8.77513 & 10.3 & $2.91^{+0.4}_{-0.3}\times10^{9}$ & $10\pm4.\times10^{40}$ & 9.1 \\
12608 & 32.92161 & -9.30599 & NGC   853 & $5.0\times10^{-3}$ & SRGe J021141.3-091817 & 32.92202 & -9.30458 & 11.2 & $3.4^{+0.5}_{-0.4}\times10^{6}$ & $3.3\pm1.0\times10^{39}$ & 454.7 \\
13693 & 336.8779 & -9.66499 & LEDA  988084 & $5.7\times10^{-3}$ & SRGe J222730.9-093959 & 336.87871 & -9.66634 & 7.2 & $2.6^{+0.7}_{-0.4}\times10^{7}$ & $8.8\pm3.1\times10^{39}$ & 13.2 \\
15429 & 206.67245 & 59.82294 & MCG+10-20-028 & $6.2\times10^{-3}$ & SRGe J134641.8+594925 & 206.67409 & 59.82353 & 8.9 & $5.1^{+1.3}_{-1.0}\times10^{8}$ & $3.7\pm1.5\times10^{39}$ & 15.2 \\
15677 & 232.06538 & 52.84361 & - & 0.019 & SRGe J152816.3+525041 & 232.06794 & 52.84466 & 8.7 & $2.6^{+0.7}_{-0.4}\times10^{8}$ & $1.9\pm1.3\times10^{40}$ & 38.1 \\
16640 & 42.39076 & 0.04646 & LEDA 1155772 & 0.186 & SRGe J024933.7+000245 & 42.39057 & 0.0458 & 11.3 & $1.3^{+1.5}_{-0.8}\times10^{8}$ & $1.22\pm0.28\times10^{43}$ & $4.5\times10^{5}$ \\
18466 & 151.82887 & 47.00639 & UGC  5451 & $2.4\times10^{-3}$ & SRGe J100718.9+470020 & 151.82873 & 47.00552 & 9.6 & $1.2^{+0.28}_{-0.2}\times10^{8}$ & $1.1\pm0.4\times10^{39}$ & 16.4 \\
18549 & 164.24457 & 50.14062 & 2MASX J10565868+5008256 & $4.6\times10^{-3}$ & SRGe J105658.8+500828 & 164.24494 & 50.14115 & 6.7 & $1.06^{+0.4}_{-0.17}\times10^{8}$ & $5.6\pm1.5\times10^{39}$ & 23.3 \\
18900 & 187.5442 & 52.76308 & LEDA 2423061 & 0.06 & SRGe J123011.3+524543 & 187.54727 & 52.76196 & 8.5 & $2.6^{+1.3}_{-0.4}\times10^{9}$ & $5.9\pm2.3\times10^{41}$ & 157.7 \\
19521 & 158.13281 & 54.40104 & Mrk   33 & $9.8\times10^{-3}$ & SRGe J103232.0+542401 & 158.13351 & 54.40027 & 8.1 & $9.5^{+4.0}_{-1.7}\times10^{8}$ & $4.2\pm0.8\times10^{40}$ & 18.6 \\
20701 & 158.5423 & 58.06363 & Mrk 1434 & $7.5\times10^{-3}$ & SRGe J103410.1+580347 & 158.54189 & 58.0631 & 5.9 & $1.0^{+0.5}_{-0.04}\times10^{7}$ & $1.5\pm0.5\times10^{40}$ & 41.0 \\
20867 & 171.68462 & 59.15545 & IC  691 & $5.5\times10^{-3}$ & SRGe J112644.4+590921 & 171.68514 & 59.15595 & 8.0 & $9.3^{+4.0}_{-1.6}\times10^{8}$ & $1.37\pm0.26\times10^{40}$ & 4.9 \\
20929 & 176.93875 & 59.88667 & MCG+10-17-072 & $4.3\times10^{-3}$ & SRGe J114745.9+595304 & 176.94125 & 59.88458 & 10.0 & $1.7^{+0.8}_{-0.4}\times10^{8}$ & $2.4\pm1.0\times10^{39}$ & 5.5 \\
20944 & 184.00491 & 59.50806 & Mrk 1468 & 0.015 & SRGe J121601.4+593024 & 184.00592 & 59.50675 & 8.4 & $2.0^{+0.7}_{-0.4}\times10^{9}$ & $2.3\pm0.5\times10^{41}$ & 62.3 \\
21021 & 186.27255 & 61.15314 & SDSS J122505.40+610911.7 & $2.3\times10^{-3}$ & SRGe J122505.3+610912 & 186.27217 & 61.15329 & 6.8 & $8.5^{+2.4}_{-0.5}\times10^{6}$ & $1.27\pm0.31\times10^{39}$ & 3.7 \\
21115 & 197.96373 & 60.24735 & UGC  8282 & 0.011 & SRGe J131151.3+601455 & 197.96392 & 60.24863 & 8.4 & $3.3^{+0.5}_{-0.4}\times10^{8}$ & $1.5\pm0.6\times10^{40}$ & 26.5 \\
22648 & 167.01215 & 53.61681 & UGC  6182 & $4.1\times10^{-3}$ & SRGe J110803.3+533700 & 167.01393 & 53.61657 & 9.1 & $1.17^{+0.26}_{-0.16}\times10^{6}$ & $4.4\pm1.6\times10^{39}$ & $1.9\times10^{3}$ \\
22933 & 183.35844 & 54.60879 & SDSS J121326.03+543631.7 & $8.0\times10^{-3}$ & SRGe J121325.9+543632 & 183.35809 & 54.60893 & 7.3 & $2.43^{+0.4}_{-0.28}\times10^{7}$ & $1.5\pm0.4\times10^{40}$ & 161.7 \\
23630 & 320.71609 & -0.99672 & SDSS J212252.00-005949.4 & 0.184 & SRGe J212252.0-005957 & 320.71659 & -0.99911 & 9.3 & $2.82^{+0.5}_{-0.32}\times10^{9}$ & $1.8\pm0.5\times10^{43}$ & 859.9 \\
25180 & 214.87833 & 51.92448 & MCG+09-23-060 & 0.029 & SRGe J141931.2+515528 & 214.88 & 51.92432 & 7.4 & $2.76^{+0.5}_{-0.35}\times10^{9}$ & $1.36\pm0.34\times10^{41}$ & 87.8 \\
25337 & 228.05357 & 47.27519 & LEDA   87385 & 0.053 & SRGe J151212.9+471635 & 228.05386 & 47.27634 & 8.7 & $1.31^{+0.4}_{-0.26}\times10^{9}$ & $2.3\pm0.8\times10^{41}$ & 3.8 \\
27238 & 227.16699 & 52.45378 & Mrk  846 & 0.011 & SRGe J150840.4+522713 & 227.16818 & 52.45371 & 6.9 & $1.90^{+0.5}_{-0.28}\times10^{9}$ & $4.7\pm1.2\times10^{40}$ & 19.5 \\
28027 & 139.34045 & 41.91099 & UGC  4904 & $5.6\times10^{-3}$ & SRGe J091721.1+415439 & 139.33812 & 41.91083 & 8.8 & $1.09^{+0.28}_{-0.12}\times10^{6}$ & $6.3\pm2.5\times10^{39}$ & $9.9\times10^{3}$ \\
30754 & 180.76497 & 56.91676 & LEDA 2551818 & 0.019 & SRGe J120303.9+565500 & 180.76619 & 56.91658 & 7.0 & $2.7^{+0.7}_{-0.5}\times10^{9}$ & $1.16\pm0.26\times10^{41}$ & 79.1 \\
30769 & 182.85503 & 58.75883 & SBSG 1209+590 & 0.011 & SRGe J121124.6+584540 & 182.85253 & 58.76102 & 9.4 & $3.9^{+0.5}_{-0.4}\times10^{8}$ & $1.7\pm0.6\times10^{40}$ & 8.5 \\
30874 & 200.21776 & 57.64159 & NGC  5109 & $7.0\times10^{-3}$ & SRGe J132053.1+573833 & 200.22106 & 57.64258 & 8.1 & $1.1^{+1.1}_{-0.7}\times10^{9}$ & $7.4\pm2.3\times10^{39}$ & 20.6 \\
30902 & 201.47424 & 57.25445 & Mrk   66 & 0.021 & SRGe J132554.2+571518 & 201.47571 & 57.25493 & 10.2 & $1.90^{+0.21}_{-0.18}\times10^{9}$ & $5.7\pm2.3\times10^{40}$ & 7.2 \\
31794 & 216.3376 & 39.53958 & UGC  9242 & $5.5\times10^{-3}$ & SRGe J142521.3+393223 & 216.33894 & 39.5397 & 9.4 & $2.9^{+0.6}_{-0.4}\times10^{8}$ & $2.8\pm1.0\times10^{39}$ & 6.7 \\
32206 & 166.28383 & 44.74646 & [BKD2008] WR 276 & 0.022 & SRGe J110508.3+444455 & 166.28462 & 44.74856 & 9.3 & $4.9^{+0.6}_{-0.7}\times10^{8}$ & $7.3\pm2.9\times10^{40}$ & 3.4 \\
32658 & 206.99294 & 39.38945 & LEDA 2148943 & $9.3\times10^{-3}$ & SRGe J134758.2+392325 & 206.99243 & 39.39034 & 8.8 & $9.4^{+2.1}_{-1.2}\times10^{7}$ & $5.4\pm2.7\times10^{39}$ & 32.5 \\
33466 & 253.33936 & 23.08278 & SDSS J165321.44+230458.0 & 0.036 & SRGe J165321.9+230500 & 253.34143 & 23.08311 & 9.8 & $3.0^{+0.7}_{-0.4}\times10^{9}$ & $1.5\pm0.7\times10^{41}$ & 38.8 \\
34306 & 177.55119 & 42.0745 & UGC  6805 & $3.9\times10^{-3}$ & SRGe J115012.4+420427 & 177.5515 & 42.07431 & 10.1 & $4.1^{+1.0}_{-0.7}\times10^{8}$ & $2.1\pm0.9\times10^{39}$ & 4.9 \\
34500 & 181.94214 & 43.12635 & NGC  4117 & $4.7\times10^{-3}$ & SRGe J120746.3+430733 & 181.94302 & 43.12581 & 9.7 & $2.2^{+0.5}_{-0.4}\times10^{9}$ & $2.7\pm1.2\times10^{39}$ & 4.9 \\
37144 & 41.71429 & -0.48567 & - & 0.044 & SRGe J024651.9-002909 & 41.71637 & -0.48585 & 8.4 & $1.3^{+2.0}_{-1.0}\times10^{9}$ & $2.0\pm0.9\times10^{41}$ & 746.5 \\
39439 & 215.28319 & 35.29539 & MCG+06-32-004 & 0.012 & SRGe J142108.6+351741 & 215.28596 & 35.29464 & 9.8 & $2.9^{+1.5}_{-0.4}\times10^{8}$ & $1.2\pm0.5\times10^{40}$ & 3.3 \\
39591 & 230.56068 & 31.47515 & Mrk  850 & $7.3\times10^{-3}$ & SRGe J152214.7+312832 & 230.56111 & 31.47566 & 6.5 & $2.2^{+1.4}_{-0.7}\times10^{7}$ & $7.5\pm1.9\times10^{39}$ & 13.7 \\
39731 & 242.73491 & 23.99648 & SDSS J161056.37+235947.3 & 0.033 & SRGe J161056.1+235950 & 242.73375 & 23.99715 & 8.0 & $1.4^{+1.7}_{-1.0}\times10^{9}$ & $1.6\pm0.5\times10^{41}$ & 527.9 \\
39846 & 49.62766 & 40.50241 & LEDA 2166728 & 0.026 & SRGe J031830.3+403004 & 49.62641 & 40.50113 & 10.4 & $1.01^{+0.18}_{-0.12}\times10^{9}$ & $9.\pm4.\times10^{40}$ & 51.8 \\
41568 & 229.95583 & 8.07313 & SDSS J151949.39+080423.2 & 0.042 & SRGe J151949.8+080428 & 229.95769 & 8.07432 & 7.9 & $1.83^{+0.4}_{-0.28}\times10^{9}$ & $2.4\pm0.9\times10^{41}$ & 111.7 \\
43461 & 137.56142 & 59.95417 & UGC  4808 & $4.4\times10^{-3}$ & SRGe J091015.0+595710 & 137.56234 & 59.95288 & 6.5 & $2.7^{+0.6}_{-0.5}\times10^{8}$ & $7.5\pm1.6\times10^{39}$ & 115.3 \\
44628 & 231.65569 & 6.99491 & 2MASS J15263736+0659417 & 0.038 & SRGe J152637.5+065942 & 231.65612 & 6.99489 & 5.0 & $2.6^{+0.9}_{-0.7}\times10^{9}$ & $5.02\pm0.26\times10^{42}$ & 777.2 \\
45689 & 313.08316 & 0.05463 & - & 0.211 & SRGe J205220.2+000319 & 313.08396 & 0.05528 & 6.7 & $5.^{+25}_{-5.}\times10^{8}$ & $9.1\pm1.9\times10^{43}$ & $8.6\times10^{5}$ \\
47785 & 203.8987 & 29.21744 & UGC  8578 & $4.1\times10^{-3}$ & SRGe J133535.6+291306 & 203.89854 & 29.21821 & 5.0 & $1.45^{+0.29}_{-0.26}\times10^{6}$ & $6.6\pm1.1\times10^{39}$ & $1.5\times10^{3}$ \\
49241 & 186.45361 & 33.54687 & NGC  4395 & $1.0\times10^{-3}$ & SRGe J122548.6+333247 & 186.45266 & 33.54625 & 7.2 & $2.5^{+0.7}_{-0.5}\times10^{7}$ & $1.04\pm0.28\times10^{39}$ & 191.9 \\
49754 & 196.48788 & 32.839 & LEDA 2014266 & 0.018 & SRGe J130557.6+325013 & 196.49004 & 32.83702 & 14.0 & $6.7^{+1.0}_{-0.7}\times10^{8}$ & $3.0\pm1.5\times10^{40}$ & 17.7 \\
50706 & 176.60869 & 34.85195 & Mrk  429 & $5.8\times10^{-3}$ & SRGe J114625.8+345105 & 176.60744 & 34.85128 & 14.8 & $2.6^{+0.6}_{-0.4}\times10^{7}$ & $3.5\pm2.0\times10^{39}$ & 53.4 \\
50842 & 205.80479 & 36.74931 & 2MASX J13431319+3644574 & 0.02 & SRGe J134313.2+364455 & 205.80521 & 36.74859 & 5.9 & $1.87^{+0.70}_{-0.34}\times10^{9}$ & $1.71\pm0.24\times10^{41}$ & 25.4 \\
50909 & 201.08405 & 36.59609 & Mrk  451 & 0.016 & SRGe J132419.9+363552 & 201.08301 & 36.59783 & 11.1 & $2.2^{+1.2}_{-0.6}\times10^{9}$ & $8.1\pm3.2\times10^{40}$ & 6.9 \\
51004 & 200.30437 & 31.22184 & UGC  8392 & 0.017 & SRGe J132113.0+311303 & 200.30419 & 31.21747 & 16.1 & $2.9^{+0.6}_{-0.4}\times10^{9}$ & $10\pm4.\times10^{40}$ & 91.8 \\
52047 & 217.16235 & 30.63451 & Z 163-59 & 0.013 & SRGe J142838.7+303809 & 217.16116 & 30.63569 & 9.7 & $1.43^{+0.4}_{-0.26}\times10^{9}$ & $5.1\pm1.7\times10^{40}$ & 33.4 \\
52077 & 217.22009 & 27.8344 & 2MASS J14285283+2750037 & 0.016 & SRGe J142852.8+275005 & 217.2198 & 27.83462 & 8.8 & $1.12^{+0.23}_{-0.14}\times10^{9}$ & $1.4\pm0.8\times10^{40}$ & 7.2 \\
52695 & 227.23254 & 25.73327 & UGC  9739 & $7.3\times10^{-3}$ & SRGe J150856.4+254359 & 227.23492 & 25.73296 & 11.3 & $7.4^{+1.2}_{-0.7}\times10^{7}$ & $5.1\pm1.8\times10^{39}$ & 24.1 \\
53431 & 241.38271 & 17.80728 & 2XMM J160531.8+174825 & 0.032 & SRGe J160531.8+174824 & 241.38247 & 17.80677 & 6.9 & $1.64^{+0.4}_{-0.31}\times10^{9}$ & $2.8\pm0.7\times10^{41}$ & 224.9 \\
57825 & 162.58224 & 48.33294 & - & 0.098 & SRGe J105019.2+481958 & 162.58002 & 48.33282 & 11.0 & $7.5^{+2.7}_{-1.2}\times10^{8}$ & $1.1\pm0.6\times10^{42}$ & 258.3 \\
63593 & 213.63451 & 17.98318 & ECO 2507 & 0.024 & SRGe J141432.4+175900 & 213.63489 & 17.98327 & 9.5 & $1.43^{+0.17}_{-0.15}\times10^{8}$ & $5.6\pm2.3\times10^{40}$ & 9.9 \\
63898 & 233.69412 & 12.44725 & LEDA 3091021 & $9.0\times10^{-3}$ & SRGe J153446.8+122654 & 233.69496 & 12.44838 & 9.1 & $9.6^{+1.2}_{-0.9}\times10^{7}$ & $1.4\pm0.4\times10^{40}$ & 37.6 \\
64321 & 233.64713 & 15.20097 & NGC  5954 & $7.5\times10^{-3}$ & SRGe J153435.1+151203 & 233.64612 & 15.20094 & 9.0 & $7.9^{+7.0}_{-3.0}\times10^{6}$ & $1.08\pm0.31\times10^{40}$ & 40.5 \\
64653 & 234.13292 & 16.44037 & UGC  9925 & 0.011 & SRGe J153632.1+162628 & 234.13383 & 16.44113 & 5.6 & $6.8^{+1.7}_{-1.2}\times10^{8}$ & $4.4\pm0.8\times10^{40}$ & 163.3 \\
65319 & 240.86063 & 19.16268 & Mrk  296 & 0.016 & SRGe J160326.5+190940 & 240.86027 & 19.16108 & 7.7 & $2.1^{+2.1}_{-1.3}\times10^{9}$ & $2.1\pm0.9\times10^{40}$ & 22.7 \\

 \caption{(1) - ID from this paper; (2,3) - SDSS fibre position in deg, (4) Galaxy name from Simbad; (5) redshift (from MPA-JHU or NED, see sect. \ref{sect:data:galaxy} for details); (6,7,8) eROSITA name, coordinates (in deg); (9) eROSITA positional error, the radius of the 98\% probability circle in arcsec ; (10) Stellar mass (from MPA-JHU, solar masses); (11)  X-ray luminosity in the 0.3--8 keV energy range (erg/s); (12) ratio of the observed X-ray luminosity to the expected X-ray binary emission (0.5--8 keV). }
\end{longtable}

}
\end{landscape}

\bsp	
\label{lastpage}
\end{document}